\documentclass[%
twocolumn,
superscriptaddress,
amsmath,amssymb,
aps,
prb,
]{revtex4-2}

\usepackage{graphicx}
\usepackage{dcolumn}
\usepackage{bm}
\usepackage{tikz}
\usepackage{float}
\usepackage{subfigure}
\usepackage[utf8]{inputenc}
\usepackage{booktabs, array, mathptmx, float, tabularx, booktabs, lipsum, amsmath,multirow,braket}
\graphicspath{{figs/}{figsgaoerb/}} 
\usepackage[colorlinks,linkcolor=blue,anchorcolor=blue,citecolor=blue]{hyperref}
\usepackage{mathtools,amsmath,graphicx,array,amssymb}
\usepackage{dcolumn}
\usepackage{bm}
\usepackage{xcolor}
\usepackage{siunitx}
\usepackage{hyperref}
\hypersetup{colorlinks=true,
	linkcolor=blue,
	filecolor=blue,
	urlcolor=blue,
	citecolor=blue,
}
\usepackage{url}
\usepackage[utf8]{inputenc}
\usepackage[english]{babel}
\usepackage{upgreek}

\begin{document}


	\title{Electronic orders on the kagome lattice at the lower Van Hove filling}


	\author{Yi-Qun Liu}
    \author{Yan-Bin Liu}
    \affiliation{National Laboratory of Solid State Microstructures $\&$ School of Physics, Nanjing University, Nanjing 210093, China}
	\author{Wan-Sheng Wang}
	\affiliation{Department of Physics, Ningbo University, Ningbo 315211, China}
    \author{Da Wang}
    \email{dawang@nju.edu.cn}

	\author{Qiang-Hua Wang}
	\email{qhwang@nju.edu.cn}
	\affiliation{National Laboratory of Solid State Microstructures $\&$ School of Physics, Nanjing University, Nanjing 210093, China}
    \affiliation{Collaborative Innovation Center of Advanced Microstructures, Nanjing University, Nanjing 210093, China}



\begin{abstract}
We study the electronic orders at the lower van Hove filling in the kagome lattice. In the weak limit of the Hubbard interaction $U$ versus the hopping parameter $t$, we find that the system develops itinerant ferromagnetism; In the intermediate range of $U$, we find the system develops noncollinear magnetic order with orthogonal spin moments on nearest-neighbor bonds. This is in fact a Chern insulator supporting quantized anomalous Hall conductance; In the strong $U$ limit, we map the Hubbard model to the $t$-$J$ model with $J = 4t^2/U$. For moderate values of $J$ we recover the noncollinear magnetic order obtained in the Hubbard model. However, in the limit of $J\to 0$ (or $U\to \infty$) we find the ferromagnetic order revives. The results are obtained by combination of the random-phase approximation and functional renormalization group in the weak to moderate limit of $U$, and the variational quantum Monte Carlo for the $t$-$J$ model in the strong coupling limit. The phase diagram is distinctly different to that at the higher van Hove filling studied earlier, and the difference can be attributed to the lack of particle-hole symmetry in the band structure with respect to the Dirac point.
\end{abstract}


	\maketitle


\section{\label{sec:level1}INTRODUCTION}
The kagome lattice has received broad research interest in recent years \cite{Yin_N_2022,Liu2020,Zhang2022,Yin2020,PhysRevB.101.094107,Lopez-Bezanilla2023} due to its several fascinating properties.
One is its high degree of geometrical frustration, which hinders the {antiferromagnetic} ordering and hence leaves possibilities of various quantum spin liquids or valence bond solid states, for the half-filled case (one electron per site) in the Mott limit (with local spins described by the Heisenberg model) \cite{PhysRevLett.101.117203,PhysRevB.76.180407,PhysRevLett.98.117205,science,PhysRevLett.118.137202,PhysRevB.92.125122,PhysRevB.91.041124,Zhou_RMP_2017,PhysRevB.105.094409,PhysRevResearch.5.L012025,prb065405,prl067201}.
Away from the Mott limit, the charge degrees of freedom cannot be neglected. Actually, even the single-orbital tight binding model already gives interesting band structures, containing a flat band at the band top and two van Hove fillings with filling fractions $1/2$ and $5/6$ per site. For all these cases, the interaction effects are expected to be strong due to the divergence of density of states arising from the van Hove singularity and the perfect Fermi surface nesting.
The upper van Hove filling $5/6$ has been intensively studied. Various electronic orders are predicted, such as the 120$^\circ$ antiferromagnetic state (AFM120$^\circ$), chiral $d+id$ superconductivity, and the charge-bond order (CBO) in the star-of-David pattern \cite{PhysRevB.85.144402,PhysRevB.87.115135,Kiesel2013,PhysRevB.99.085119}. In particular, the CBO state may be relevant in the new kagome superconductors AV$_{3}$Sb$_{5}$ (A=K, Cs, Rb) \cite{kagome.material,PhysRevLett.125.247002, PhysRevMaterials.5.034801, Yin_2021}, although there is still a debate on whether the CBO state is chiral in the material.

\begin{figure*}
	\begin{tikzpicture}
	\node[anchor=south west,inner sep=0] (image) at (0,0) {\subfigure{\includegraphics[trim=120 45 130 160,clip,width=0.7\textwidth]{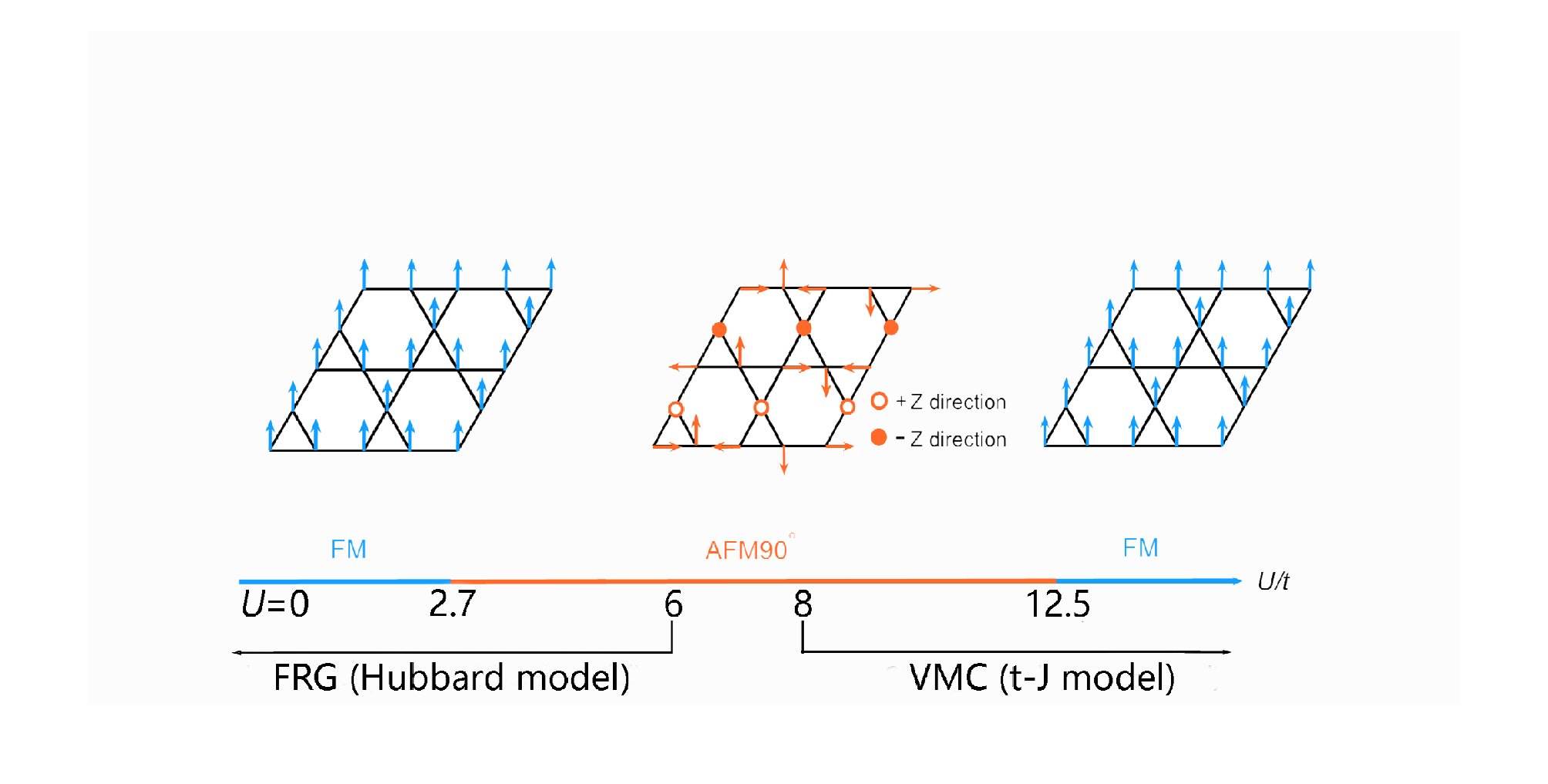}}};
	\end{tikzpicture}
	\caption {The phase diagram of the Hubbard model on the kagome lattice at the lower van Hove filling level. The magnetic ordering configurations for the FM and AFM90$^{\circ}$ states are sketched, respectively.}
	\label{fig:finalphase}
\end{figure*}

To our knowledge, {the lower van Hove filling level $1/2$ has rarely been studied in the literature \cite{PhysRevB.90.245119}.}
Note that the shape of the Fermi surface at this level is the same as that at the upper van Hove filling $5/6$. The two levels are symmetric with respect to the Dirac point in the band dispersion, similarly to that in the honeycomb lattice. However, the upper van Hove level is closer to the flat band on the top, hence the two levels are intrinsically different. In other words, the band structure lacks particle-hole symmetry (as that in the honeycomb lattice) with respect to Dirac point.
For this reason, in this work, we focus on the lower van Hove filling to investigate its electronic instabilities caused by the Hubbard repulsion $U$.
For small and intermediate $U$, we perform random phase approximation (RPA) and functional renormalization group (FRG) calculations. As $U$ increases, we find a transition from the ferromagnetic (FM) order to the noncollinear antiferromagnetic order with orthogonal nearest spin moments (AFM90$^{\circ}$), in contrast to the AFM120$^{\circ}$ at the upper van Hove filling. The AFM90$^{\circ}$ state is a Chern insulator supporting quantized anomalous Hall conductance.
For even stronger $U$, we perform variational Monte Carlo (VMC) to study the effective $t$-$J$ model. In the overlapping regime of $U$ we confirm the AFM90$^{\circ}$ order. However, in the extreme limit of large $U$ (or small $J$) we find a first order transition back to the FM state. The results are summarized as a ground state phase diagram in Fig.~\ref{fig:finalphase}.  We should note that the magnetic states can order in the zero temperature limit, while at finite temperatures they should be understood as signifying the relevant spin correlations, in view of the Mermin-Wagner theorem for spin orderings in the SU(2) symmetric two-dimensional system.

 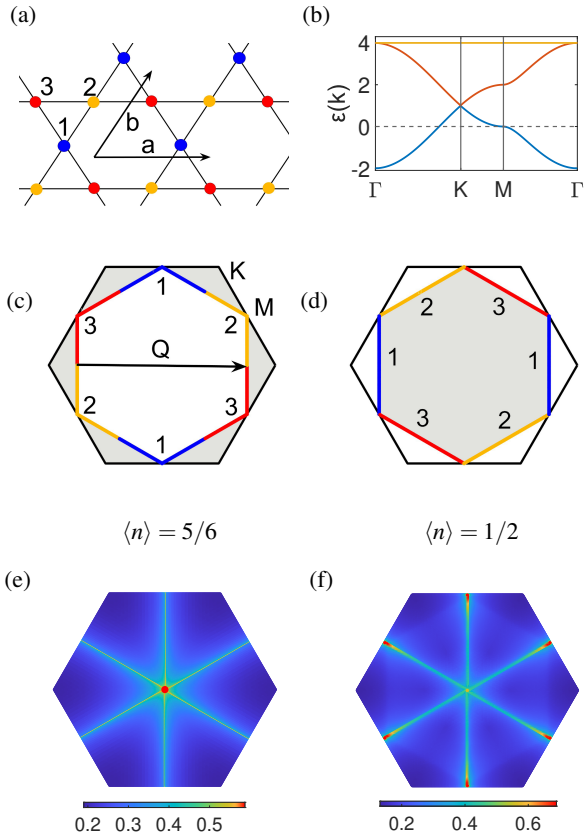
\begin{figure}
    	\centering
    	  \begin{tikzpicture}
    		\node[anchor=south west,inner sep=0] (image) at (0,0) {\subfigure{\includegraphics[trim=50 0 208 0 ,clip,width=0.2\textwidth]{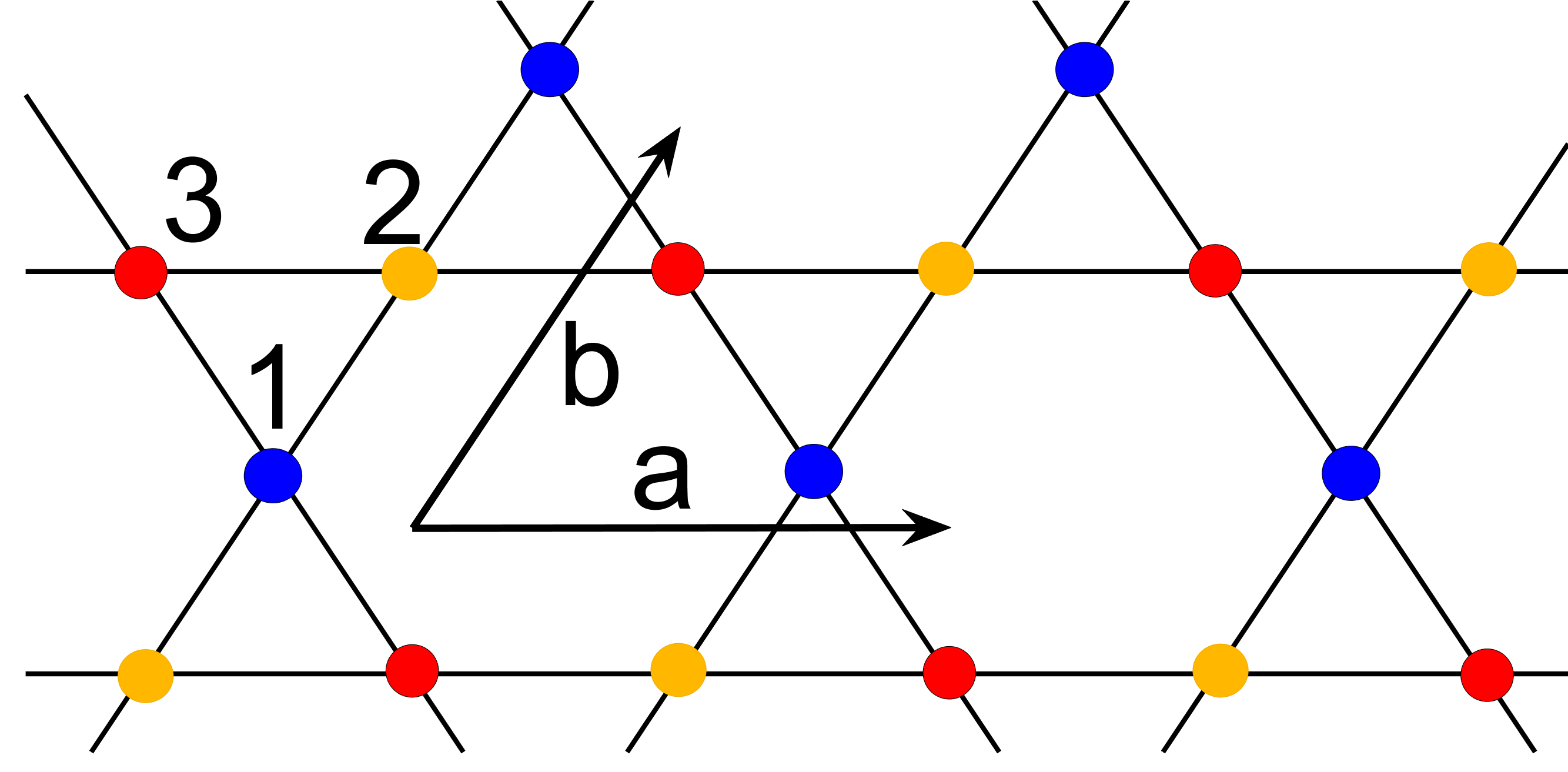}\label{fig:lattice}}};
    		\node[above left] at (0.35,2.5) {(a)};
    	\end{tikzpicture}
    	\begin{tikzpicture}

    		\node[anchor=south west,inner sep=0] (image) at (0,0) {\subfigure{\includegraphics[trim=40 10 346 20,clip,width=0.2\textwidth]{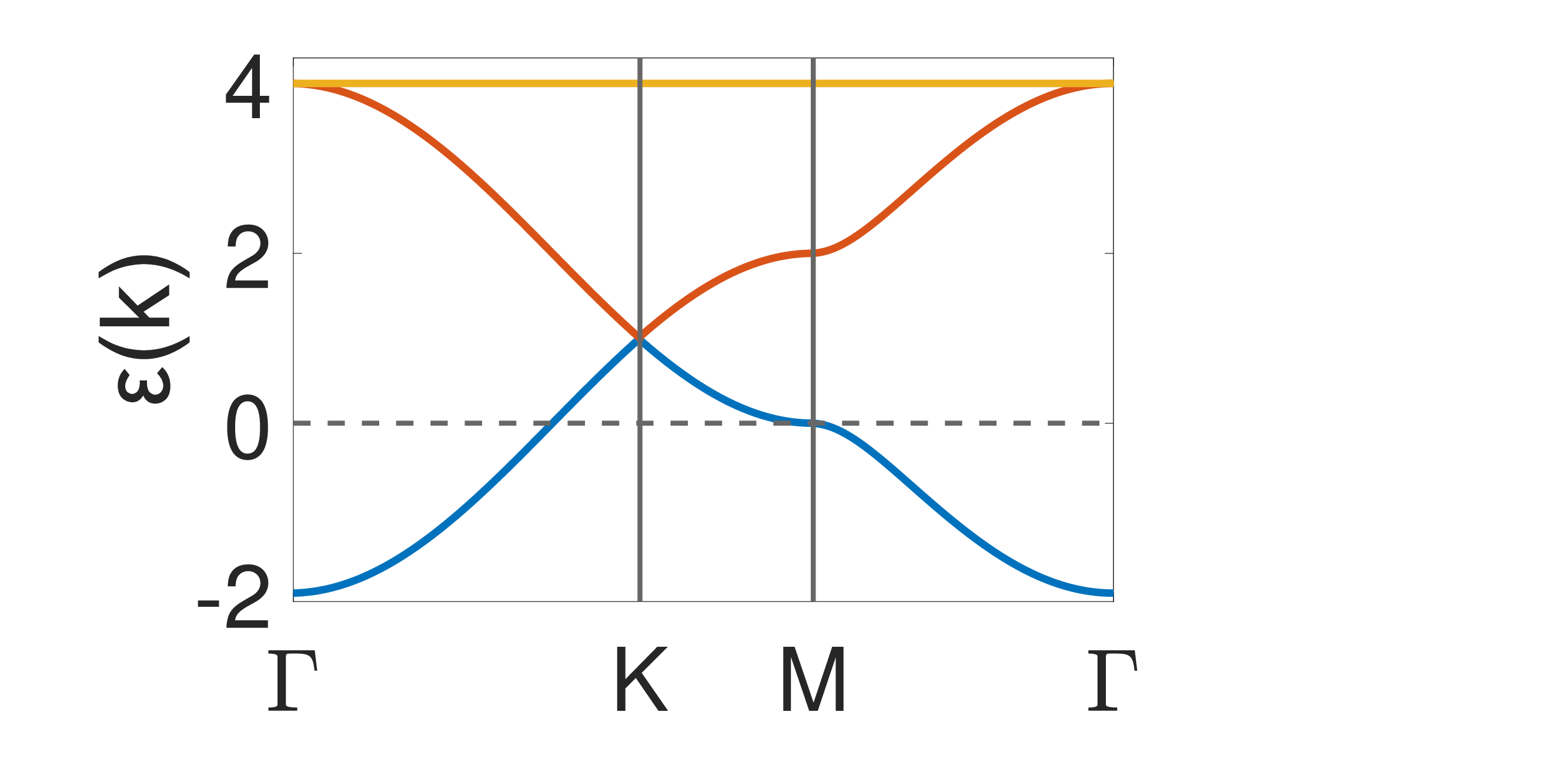}\label{fig:band}}};
    		\node[above left] at (0.32,2.5) {(b)};
    	\end{tikzpicture}
    	  	\begin{tikzpicture}
    	  		\hspace{-2pt}
    		\node[anchor=south west,inner sep=0] (image) at (0,0) {\subfigure{\includegraphics[trim=280 0 260 0,clip,width=0.18\textwidth]{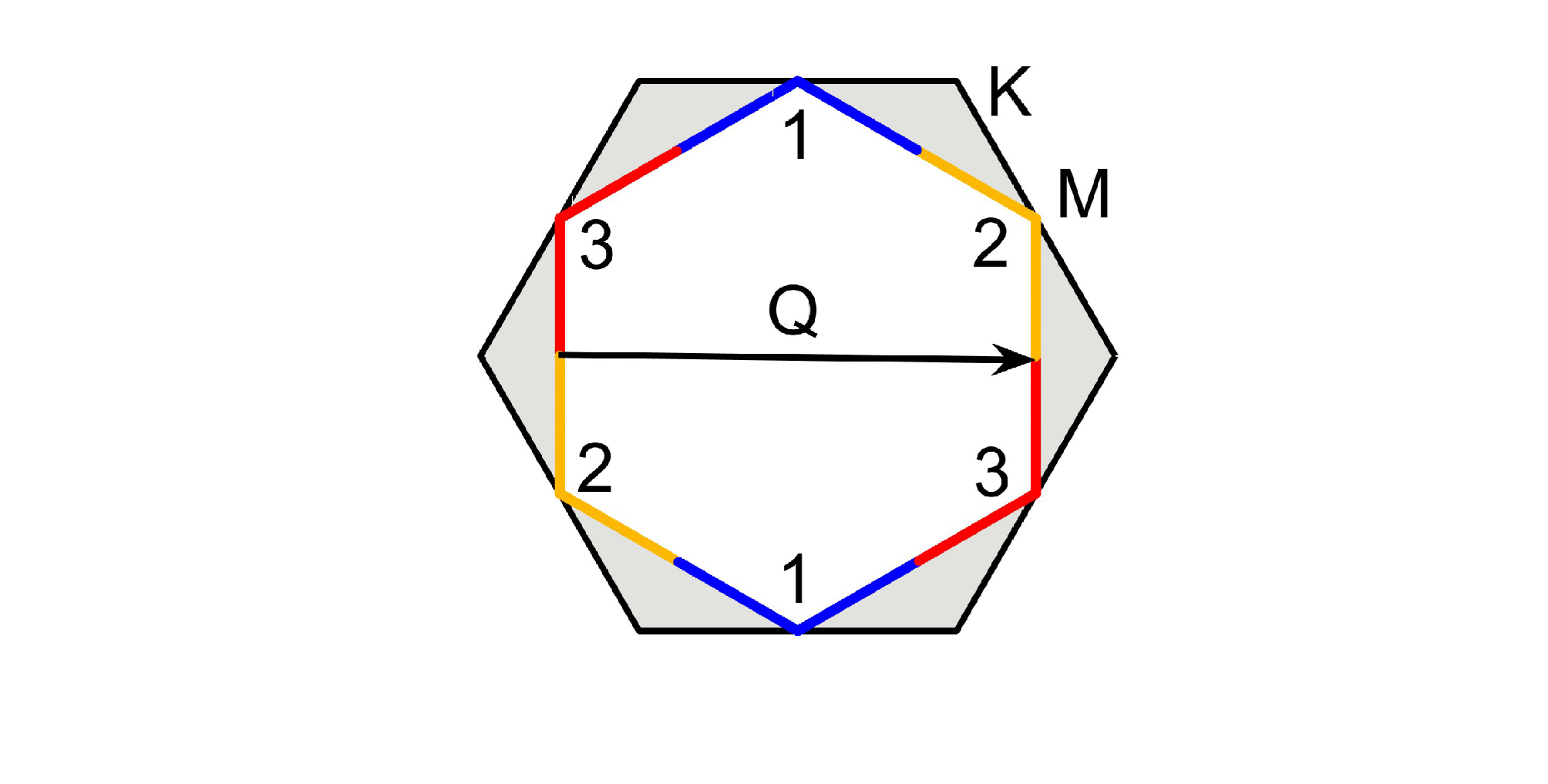}\label{fig:umee}}};
    		\node[above left] at (0.05,2.8) {(c)};
    		\node[above left] at (2.5,-0.3) {$\braket{n}=5/6$};
    	\end{tikzpicture}
    	\begin{tikzpicture}
    		\hspace{0pt}
    		\node[anchor=south west,inner sep=0] (image) at (0,0) {\subfigure{\includegraphics[trim=280 0 260 0,clip,width=0.18\textwidth]{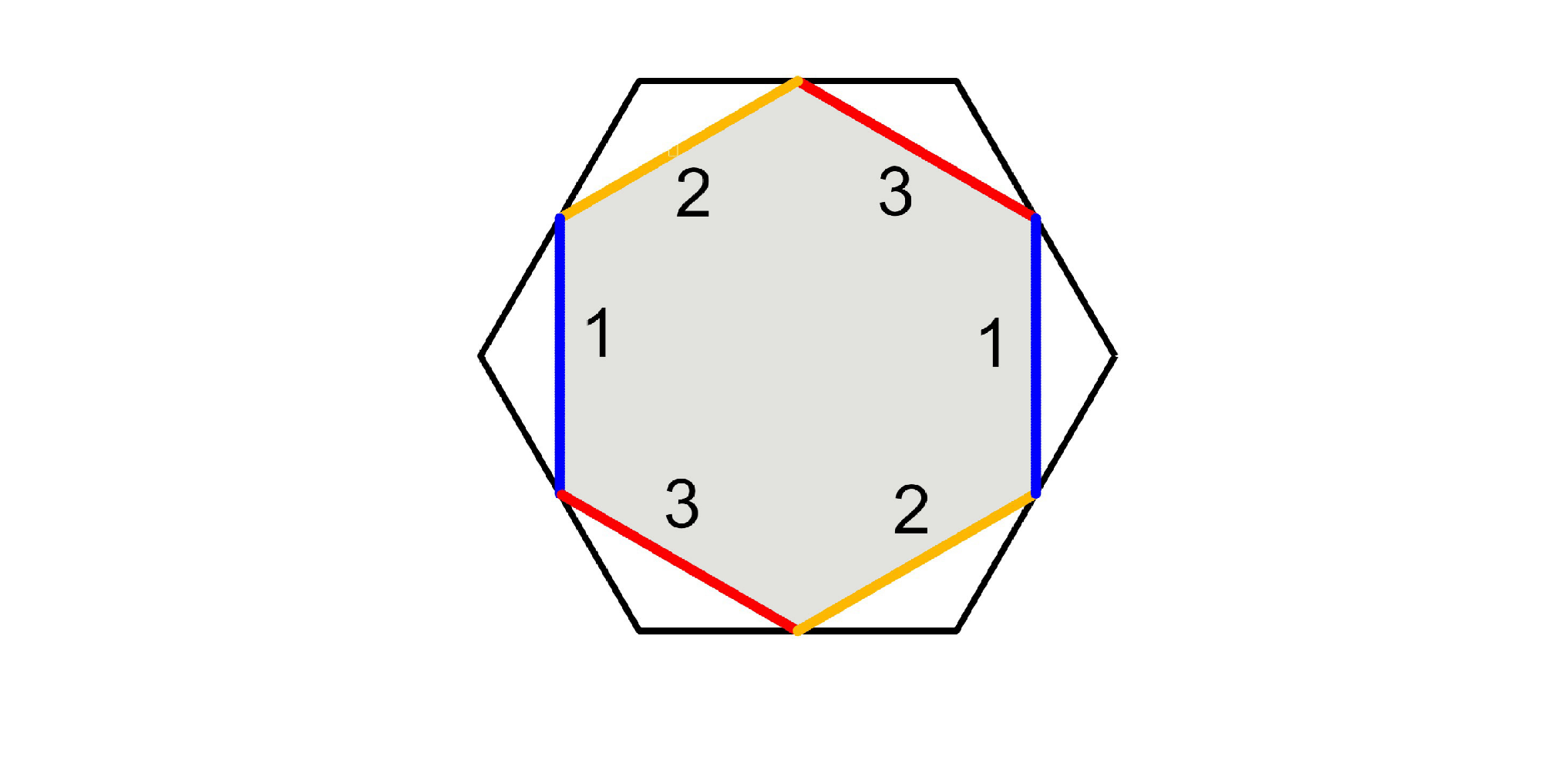}\label{fig:lmee}}};
    		\node[above left] at (-0.04,2.8) {(d)};
    		\node[above left] at (2.5,-0.3) {$\braket{n}=1/2$};
    	\end{tikzpicture}
  	  	\begin{tikzpicture}
  		\hspace{-2pt}
  		\node[anchor=south west,inner sep=0] (image) at (0,0) {\subfigure{\includegraphics[trim=370 30 370 0,clip,width=0.18\textwidth]{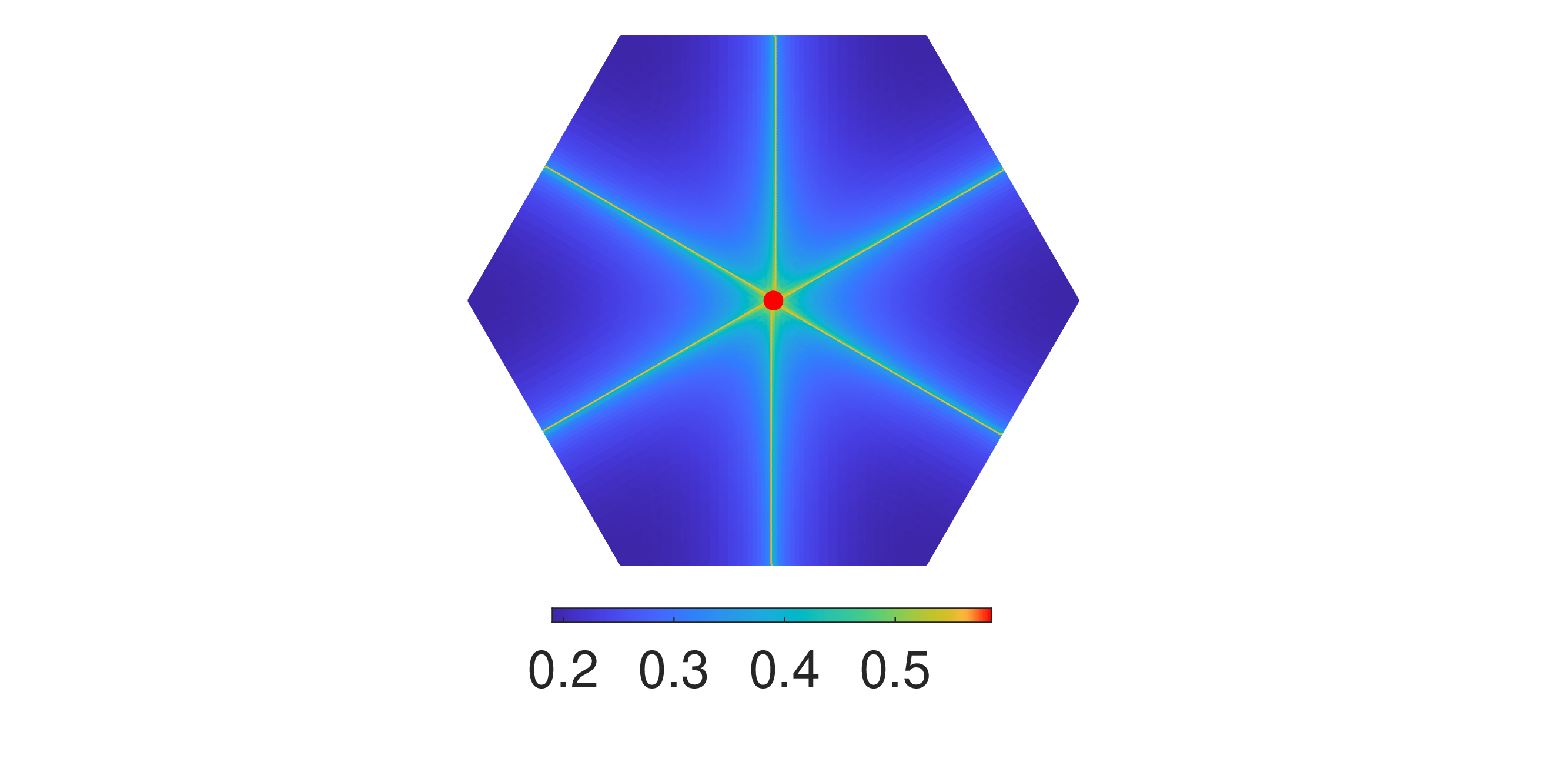}\label{fig:u54}}};
  		\node[above left] at (-0.1,3.5) {(e)};
  	\end{tikzpicture}
  	\begin{tikzpicture}
  		\hspace{0pt}
  		\node[anchor=south west,inner sep=0] (image) at (0,0) {\subfigure{\includegraphics[trim=380 30 380 0,clip,width=0.18\textwidth]{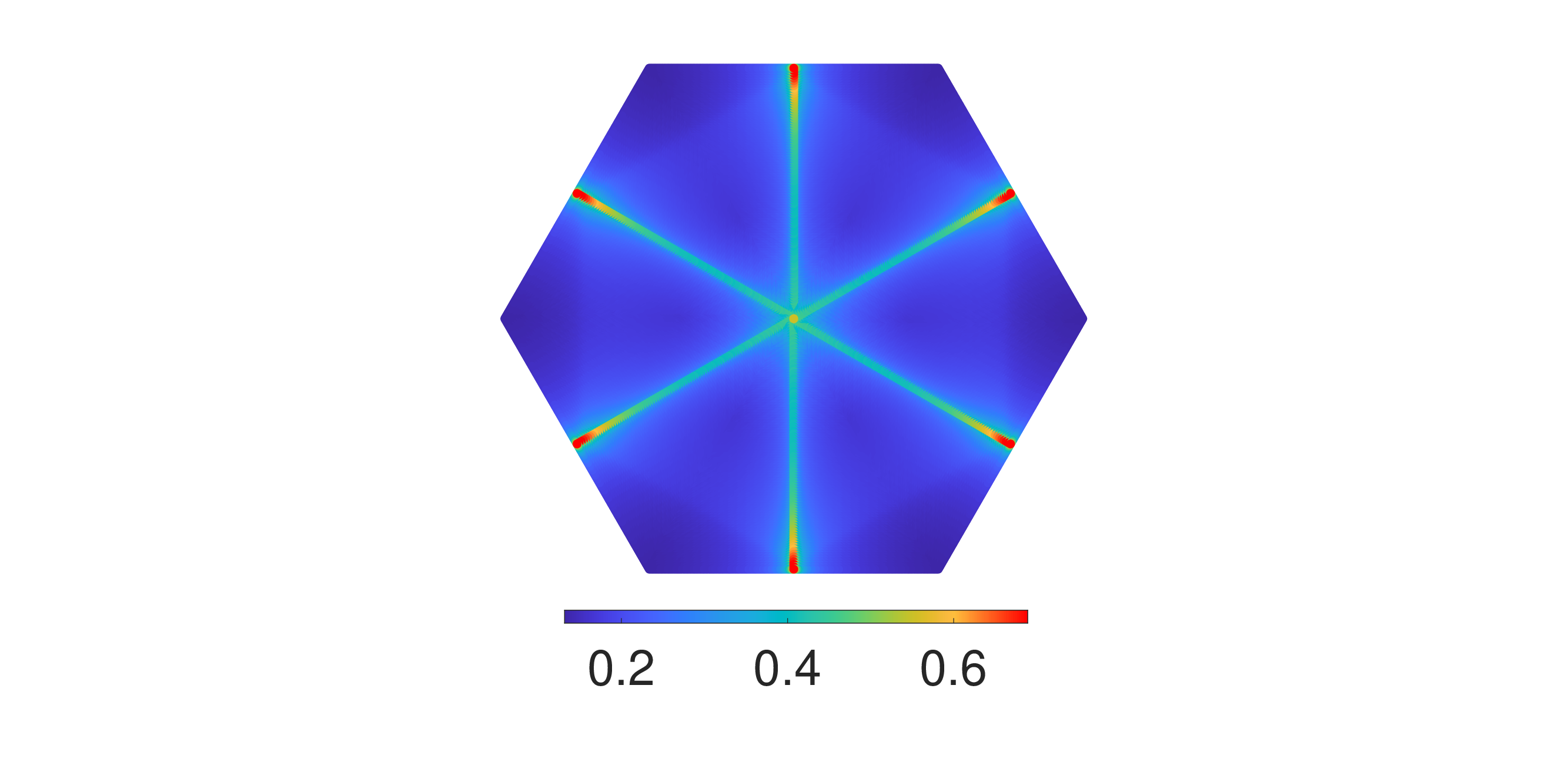}\label{fig:l54}}};
  		\node[above left] at (0,3.5) {(f)};
  	\end{tikzpicture}
    	\caption{ (a) Kagome lattice with $\mathbf{a},\mathbf{b}$ two primitive translation vectors and $1,2,3$ denoting three sublattices. (b) Band dispersion along the high-symmetry line. (c) Fermi surface at the upper Van Hove filling level. The line color indicates the leading sublattice component, and $\mathbf{Q}$ is one nesting vector. {The gray area denotes the occupied states in the conduction band.} (d) is similar to (c) but at the lower Van Hove filling level. (e) and (f) plot leading eigenvalues of the bare susceptibility matrix $\chi_0(\mathbf{q})$ at $T=0.0005t$ for upper and lower van Hove filling levels, respectively.}
    	\label{fig:1}
    \end{figure}

\section{\label{sec:level2}MODEL}
The single-orbital Hubbard model on the kagome lattice is described by the Hamiltonian
\begin{equation}
H = -t\sum_{\braket{ij}\sigma}(c^{\dagger}_{i\sigma}c_{j\sigma}+{\rm H.c.}) + U\sum_{i}n_{i\uparrow} n_{i\downarrow} - \mu \sum_{i\sigma}n_{i\sigma},
\end{equation}
where the operator $c^\dagger_{i\sigma}$ creates an electron with spin $\sigma$ on site $i$, and $n_{i\sigma} = c^{\dagger}_{i\sigma} c_{i\sigma}$.
The parameter $t$ is the hopping integral between the nearest neighbor sites $\braket{ij}$, $U$ is the onsite Hubbard repulsion, $\mu$ is the chemical potential to fix the system at the lower van Hove filling.
The kagome lattice is shown in Fig.~\ref{fig:lattice}. It has three sublattices in a unit cell which are labeled by different numbers, and colors. $\mathbf{a}$ and $\mathbf{b}$ are two primitive vectors.
Fig.~\ref{fig:band} plots the energy band dispersion along the high symmetry path $\Gamma-K-M-\Gamma$. The highest band is a flat band. For the other two bands, there are two van Hove fillings ($2/3\pm1/6$) when the Fermi surface touches the zone boundary $M$.
In this work we focus on the lower van Hove filling $1/2$, which corresponds to the chemical potential $\mu=-2t$.

The Fermi surface at the upper and lower van Hove fillings are plotted in Fig.~\ref{fig:umee} and Fig.\ref{fig:lmee}, respectively. They have the same shape with a nesting vector $Q$ (arrow), but the Bloch states connected by this vector are quite different, see the colored Fermi segments. It is clear that the nesting vector connects almost orthogonal Bloch states on the Fermi surface for the upper van Hove level, known as the matrix element effect \cite{PhysRevB.86.121105,PhysRevB.87.115135} that weakens the magnetic ordering arising from the onsite Hubbard interaction. Quite differently, in the case of the lower van Hove filling,  the states connected by $Q$ have the same leading sublattice components. Therefore, the nesting effect is expected to be stronger. Moreover, the three nesting vectors connects three types of Bloch states with leading components on three different sublattices, implying that the spin moments on the three sublattices may carry different wave vectors in a potential magnetic state. This may be understood as a new variety of the matrix-element effect in the kagome lattice.

To have a better idea of the matrix element effect, we calculate the bare spin susceptibility $\chi_0(\mathbf{q})$ at zero frequency,
\begin{equation}
\chi_0^{\alpha\beta}(\mathbf{q}) = -\frac{T}{N}\sum_{\mathbf{k},m}G^{\alpha\beta}(\mathbf{k},i\omega_m)G^{\beta\alpha}(\mathbf{k}+\mathbf{q},i\omega_m),
\end{equation}
where $\alpha,\beta=1,2,3$ denote sublattices, $T$ is the temperature, $N$ is the number of unit cells, $\mathbf{k}$ and $\mathbf{q}$ are momenta, $\omega_m$ is the fermionic Matsubara frequency, $G(\mathbf{k},i\omega_m)$ is the normal state Green's function (in the sublattice basis).
The largest eigenvalue of $\chi_0(\mathbf{q})$ is plotted versus $\mathbf{q}$ at an intermediate temperature $T=0.0005t$ in Fig.~\ref{fig:u54} and Fig.~\ref{fig:l54} for upper and lower van Hove fillings, respectively.
Different from the upper one with a peak at $\Gamma$, the lower one exhibits highest peaks at $M$ points corresponding to the nesting vector.
This difference implies different ground states induced by the onsite Hubbard $U$ (see below).
Of course, at very low temperature (not shown), both upper and lower van Hove fillings have the highest peak at $\mathbf{q}=0$, which directly comes from the logarithmic divergence of the density of states at these two van Hove singularities, implying the Stoner FM instability in the small $U$ limit.
In the following, we will combine both weak and strong coupling methods to analyze the ground states for the Hubbard model at the lower van Hove filling.

\section{RPA analysis}

\begin{figure}
		\hspace{-27pt}
\includegraphics[trim=50 0 0 0,clip,width=0.44\textwidth]{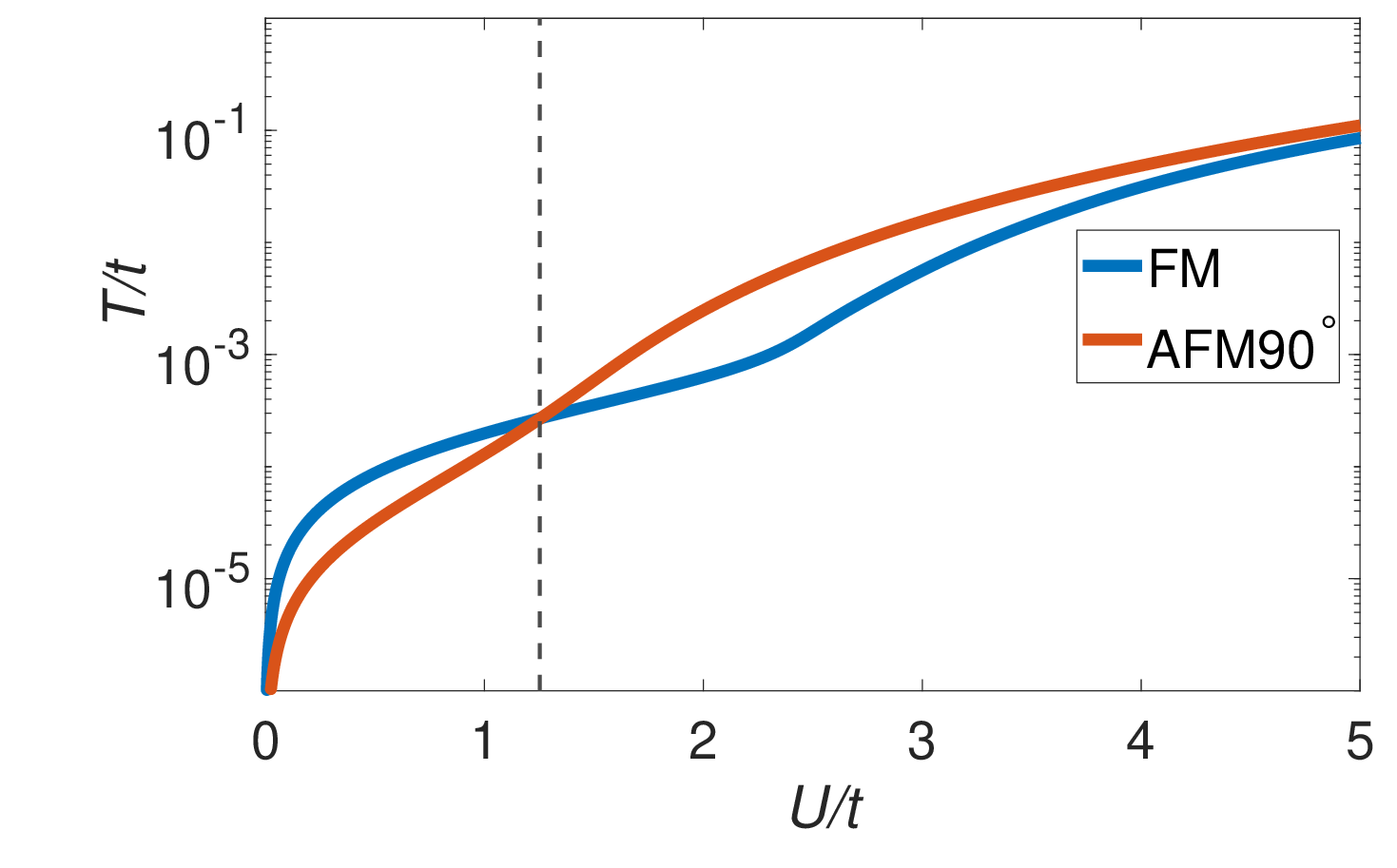}
\caption{The Stoner instability temperature versus Hubbard $U$ for FM (blue line) and AFM90$^{\circ}$ (red line) orders, respectively.}
\label{fig:rpa}
\end{figure}

\begin{figure}
		\hspace{-16pt}
	\includegraphics[trim=15 20 30 230,clip,width=0.45\textwidth]{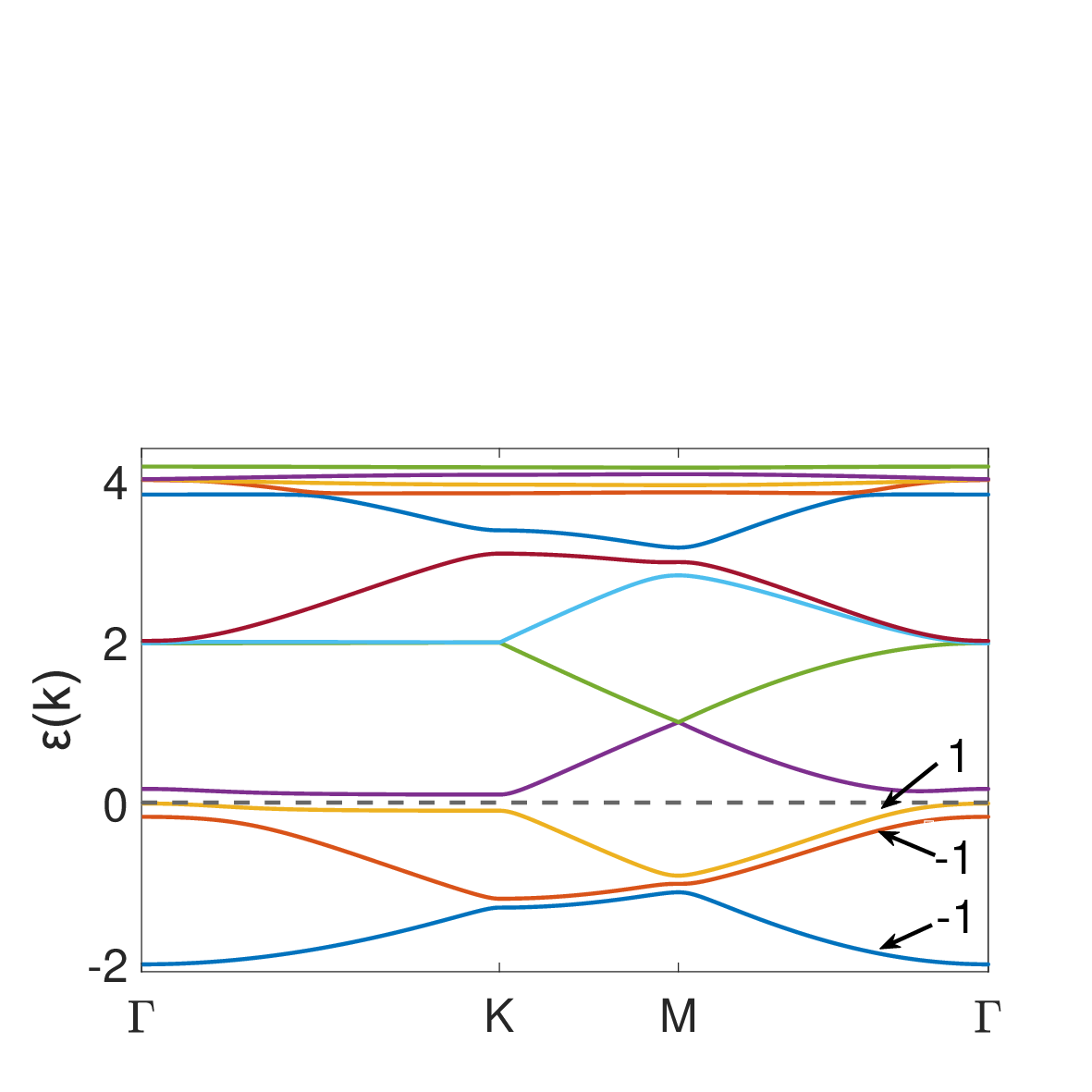}
	\caption{Band dispersion (with two-fold band degeneracy) in the AFM90$^{\circ}$ state. The Chern numbers of the lowest three bands are shown in the figure.}
	\label{fig:chernnumber}
\end{figure}

First, we perform RPA to get a taste of the magnetic instabilities, which should be reliable at least in the weak coupling limit\cite{PhysRevB.99.085119,PhysRevB.94.165141}. The spin susceptibility matrix $\chi(\mathbf{q})$ is renormalized by the onsite Hubbard $U$ (diagonal in sublattice-space) as
\begin{equation}
	\chi(\mathbf{q}) = \chi_0(\mathbf{q})[1-U\chi_0(\mathbf{q})]^{-1} .
\end{equation}
As the temperature is lowered, the first magnetic instability is signaled by the first divergence of the susceptibility. This occurs at $U=1/\lambda(\mathbf{q})$, where $\lambda(\mathbf{q})$ is the largest eigenvalue of $\chi_0(\mathbf{q})$ (and is temperature dependent). This is just the Stoner criterion extended for a multi-sublattice system. The corresponding spin ordering configuration is determined by the eigenvector of $\chi_0(\mathbf{q})$.
In Fig.~\ref{fig:rpa}, we plot the transition temperature $T_c$ determined by the Stoner criterion as a function of the Hubbard interaction $U$ for $\mathbf{q}$ at $\Gamma$ and M, respectively.
For small $U$, $T_c(\Gamma)>T_c(M)$, indicating the system enters a magnetic state {with the spin configuration keeping the same} from unitcell to unitcell. Furthermore, the eigenvector is $[1,1,1]/\sqrt{3}$, so the moment is identical on the three sublattices, and it is just a conventional FM state.
For larger $U$, $T_c(M)>T_c(\Gamma)$, indicating the system develops a magnetic order with momentum $M$. By symmetry, there are three degenerate $M$'s, and the eivenvectors at these momenta are $[1,0,0]$, $[0,1,0]$, and $[0,0,1]$, respectively. Given the degeneracy, it is necessary to perform a mean field calculation to determine the ordered state, which can combine the degenerate modes optimally to minimize the (free) energy.
The mean field calculation is straightforward, and yields a novel noncollinear antiferromagnetic order with orthogonal nearest spin moments, as shown in Fig.~\ref{fig:finalphase}, which we dub as AFM90$^{\circ}$. {Details of the mean field calculation can be found in appendix \ref{MFTdetails}.}
Interestingly, in this state, a finite energy gap opens at the Fermi level, and because of the solid angle swept by the spin moments on the three sublattices, the Chern numbers for the lowest three bands are $C=\pm1$ as shown in Fig.~\ref{fig:chernnumber}. Therefore, The total Chern number of the filled bands is $-2$ ($2$ from the two-fold band degeneracy).
In other words, we obtain a Chern insulator supporting quantized anomalous Hall conductance \cite{PhysRevB.90.245119}.
Here, in order to calculate the Chern number for each two degenerate bands $(n,n+1)$, we use the projecting operator $P_{n,n+1}(k_x,k_y)=\sum_{i=n,n+1}|ik_xk_y\rangle \langle ik_xk_y|$ to construct the Wilson loop $W_{n,n+1}(k_y)=\prod_{k_x}P_{n,n+1}(k_x,k_y)$ for each $k_y$. Then, from its two nonzero eigenvalues $e^{i\theta_{1,2}(k_y)}$, we obtain the Chern number $C_{n,n+1}=\frac{1}{2\pi}\int[\theta_1(k_y)+\theta_2(k_y)]dk_y$.

If we press on to even larger value of $U$, the FM and AFM90$^{\circ}$ transition lines tend to merge together. But at this stage, the RPA is {not reliable} to determine the correct ordered state anymore. We will come back to this strong coupling limit later when we use effective theory and variational quantum Monte Carlo to determine the ground state.

\section{Functional renormalization group}
In the above RPA, we only consider spin susceptibility, giving onsite magnetic orderings. However, the RPA is a biased theory excluding other ordering channels, such as superconductivity, site-local charge densithy wave, and charge bond order. To treat these channels on equal footing, we resort to singular-mode FRG (SMFRG) \cite{Wang_PRB_2012,PhysRevB.79.195125}.
The basic idea of FRG is to obtain the flow of one-particle-irreducible vertices as a function of the infrared cutoff energy scale $\Lambda$, which we choose as the lowest fermionic Matsubara frequency here. As in usual practice, we only consider the flow of the four-point vertices $\Gamma_{1234}$ in the effective interaction Hamiltonian
\begin{equation} H_I=\frac{1}{2} \sum_{1,2,3,4,\sigma,\sigma'}c_{1\sigma}^\dagger c_{2\sigma'}^\dagger \Gamma_{1234}c_{3\sigma'} c_{4\sigma}.\end{equation}
Here the numerical indices label the single particle state (such as the momentum/position and sublattice). The truncation at the four-point vertices is reasonable if the instability occurs at low energy scales, since the higher order vertices are irrelevant by RG dimension counting. In SMFRG, $\Gamma_{1234}$ is decomposed as scattering matrices between fermion bilinears in three Mandelstam channels: $\Gamma_{1234}=P_{12,43}=C_{13,42}=D_{14,32}$. As the energy scale is lowered, the interaction vertices flow as
\begin{align} \label{eq:flow}
\frac{\partial\Gamma_{1234}}{\partial\Lambda} &= [P\chi_{pp}P]_{12;43}+[C\chi_{ph}C]_{13;42} \nonumber\\
&+ [D\chi_{ph}C+C\chi_{ph}D-2D\chi_{ph}D]_{14;32},
\end{align}
where $\chi_{pp,ph}$ are scale-dependent one-loop functions (matrix in the bilinear base) in the particle-particle and particle-hole channels. Matrix convolution is understood on the right-hand side. After each infinitesimal integration in $\Lambda$, $\Gamma$ is rewinded into $P$, $C$ and $D$ for the right-hand-side of the flow equation. In practice, it is necessary to truncate the internal distance between two fermions in a fermion bilinear, and in our case we checked that the results converge already when we truncate the bilinears within the set of onsite and nearest-neighbor ones.
The effective interactions in superconducting (SC), spin-density-wave (SDW) and charge-density-wave (CDW) channels are given by $V_{\rm{SC}} = P$, $V_{\rm{SDW}} = -C$, and $V_{\rm{CDW}} = 2D-C$, respectively. The eigen modes of the scattering matrices are labeled by the scattering momentum of the bilinears.
The leading negative eigenvalue $S$ of these effective interactions are monitored during the SMFRG flow until the largest one diverges at a critical scale $\Lambda_c$, indicating an instability with the ordering pattern described by the corresponding eigenfunction. More technical details can be found in Ref.~\cite{PhysRevB.87.115135,Wang_PRB_2012,Xiang_PRB_2012,Wang_PRL_2019,Yang_PRB_2022,Yang_PRB_2023}.

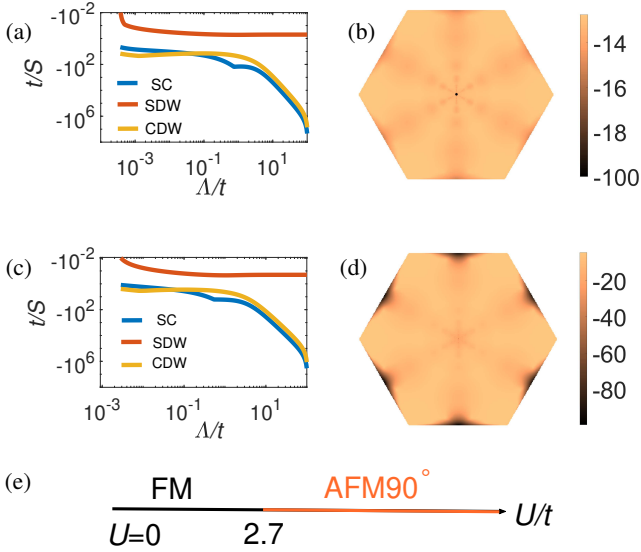
\begin{figure}
	\centering
	\begin{tikzpicture}
		\hspace{-12pt}
		\node[anchor=south west,inner sep=0] (image) at (0,0) {\subfigure{\includegraphics[trim=130 155 30 120,clip,width=0.22\textwidth]{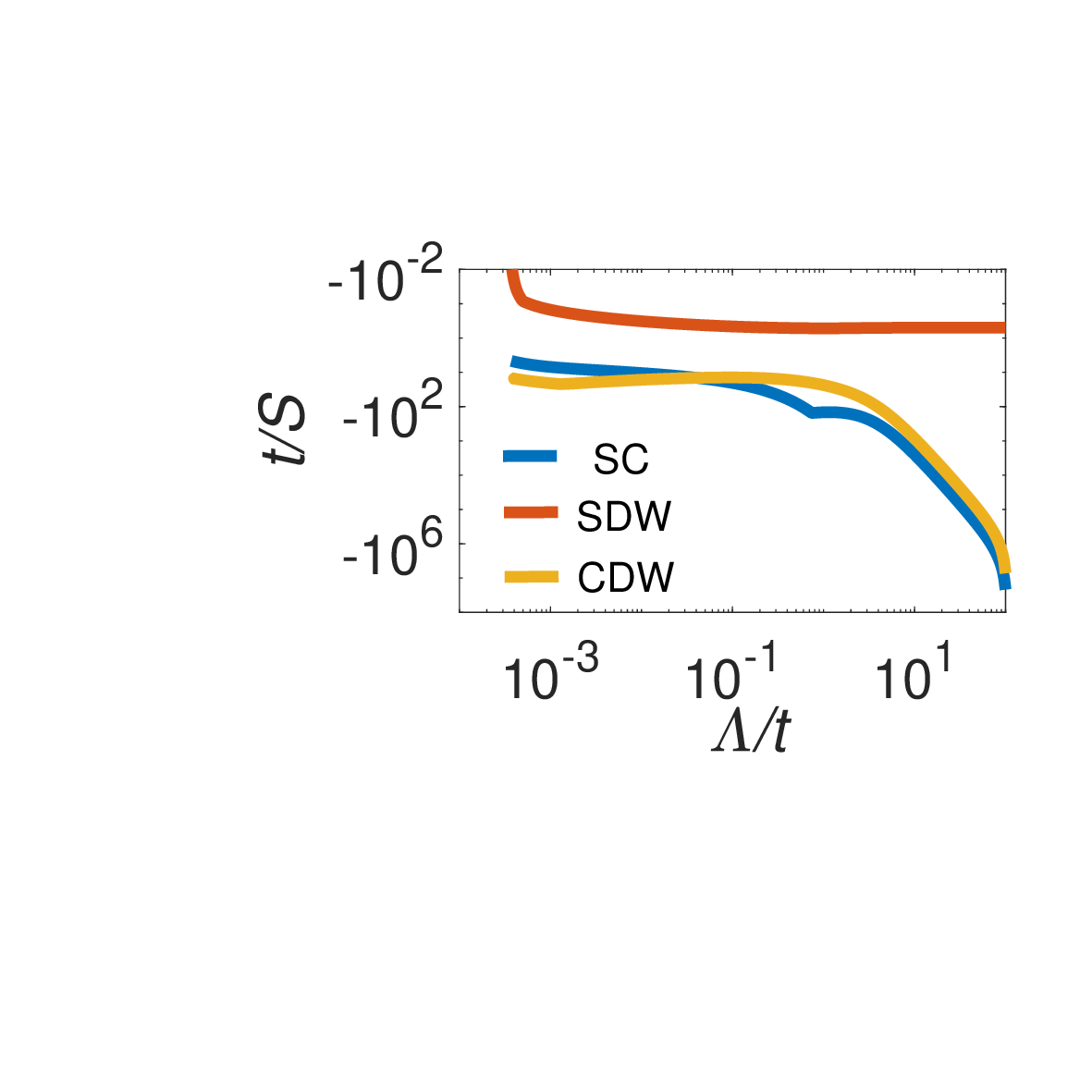}\label{fig:flow2}}};
		\node[above left] at (0.26,2.3) {(a)};
	\end{tikzpicture}
	\begin{tikzpicture}
		\hspace{-6pt}
		\node[anchor=south west,inner sep=0] (image) at (0,0) {\subfigure{\includegraphics[trim=370 0 130 40,clip,width=0.21\textwidth]{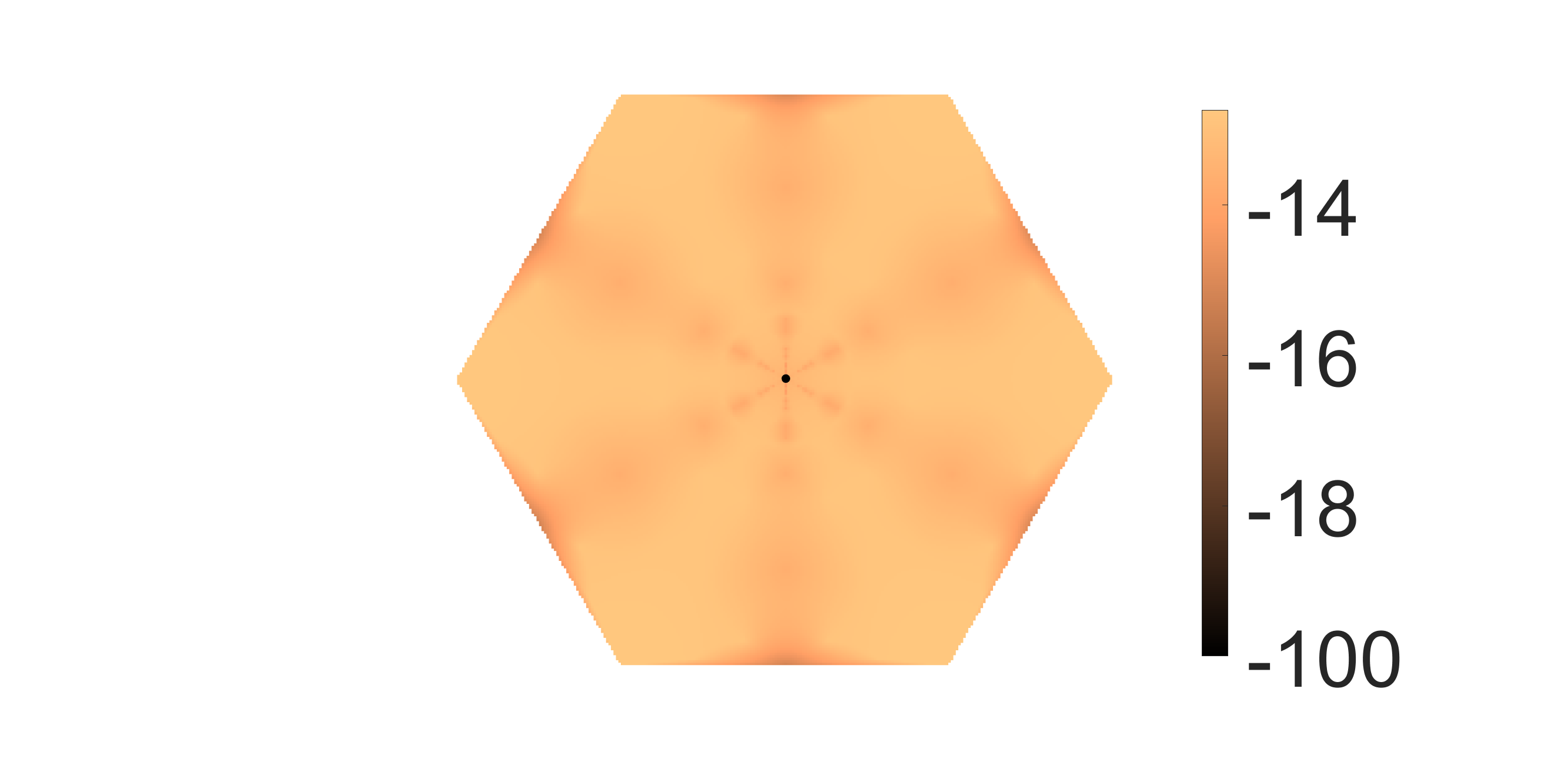}\label{fig:rev2}}};
		\node[above left] at (0.35,2.3) {(b)};
	\end{tikzpicture}
		\begin{tikzpicture}
		\hspace{-11pt}
		\node[anchor=south west,inner sep=0] (image) at (0,0) {\subfigure{\includegraphics[trim=130 120 30 180,clip,width=0.22\textwidth]{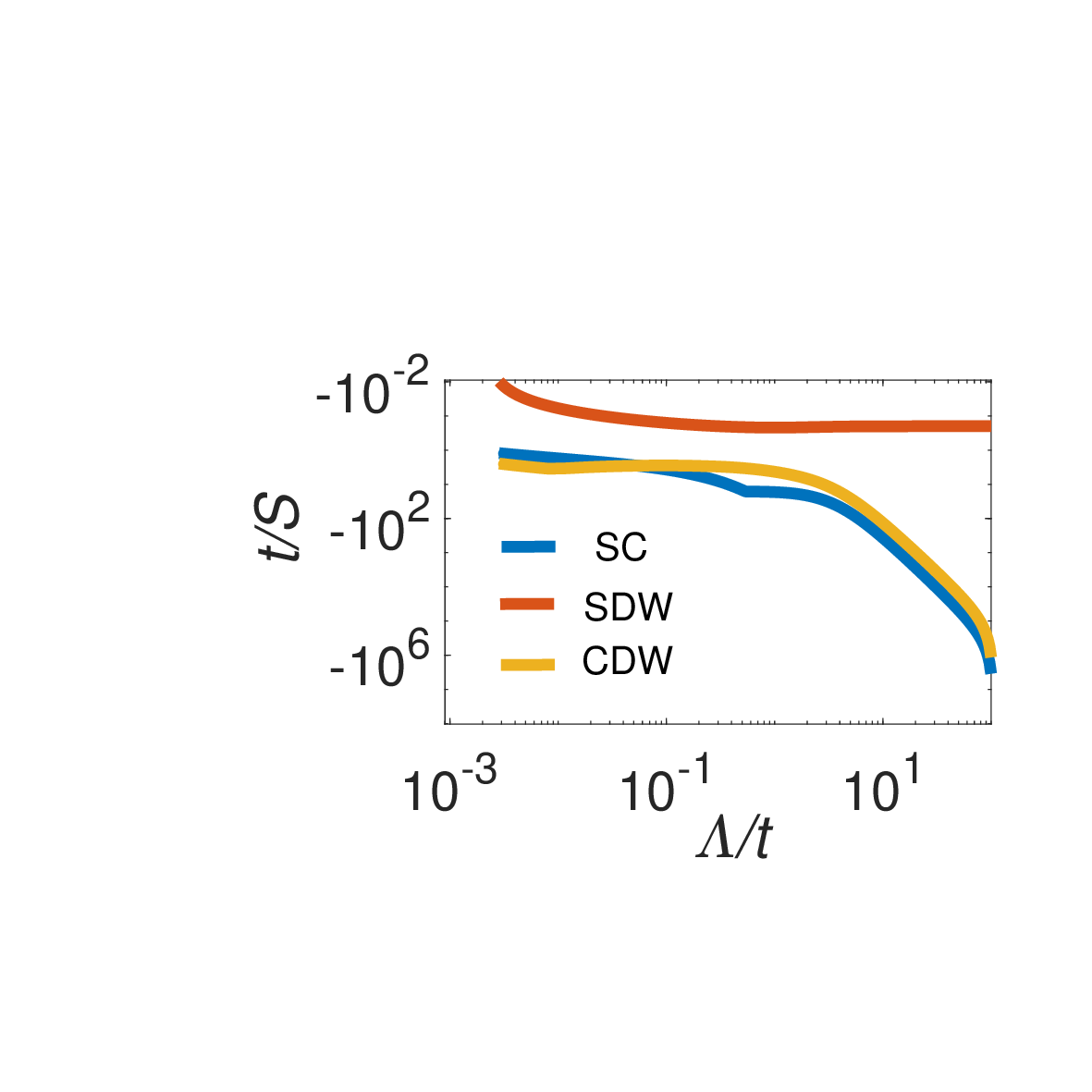}\label{fig:flow5}}};
		\node[above left] at (0.19,2.2) {(c)};
	\end{tikzpicture}
	\begin{tikzpicture}
		!\hspace{-10pt}
		\node[anchor=south west,inner sep=0] (image) at (0,0) {\subfigure{\includegraphics[trim=330 35 135 20,clip,width=0.21\textwidth]{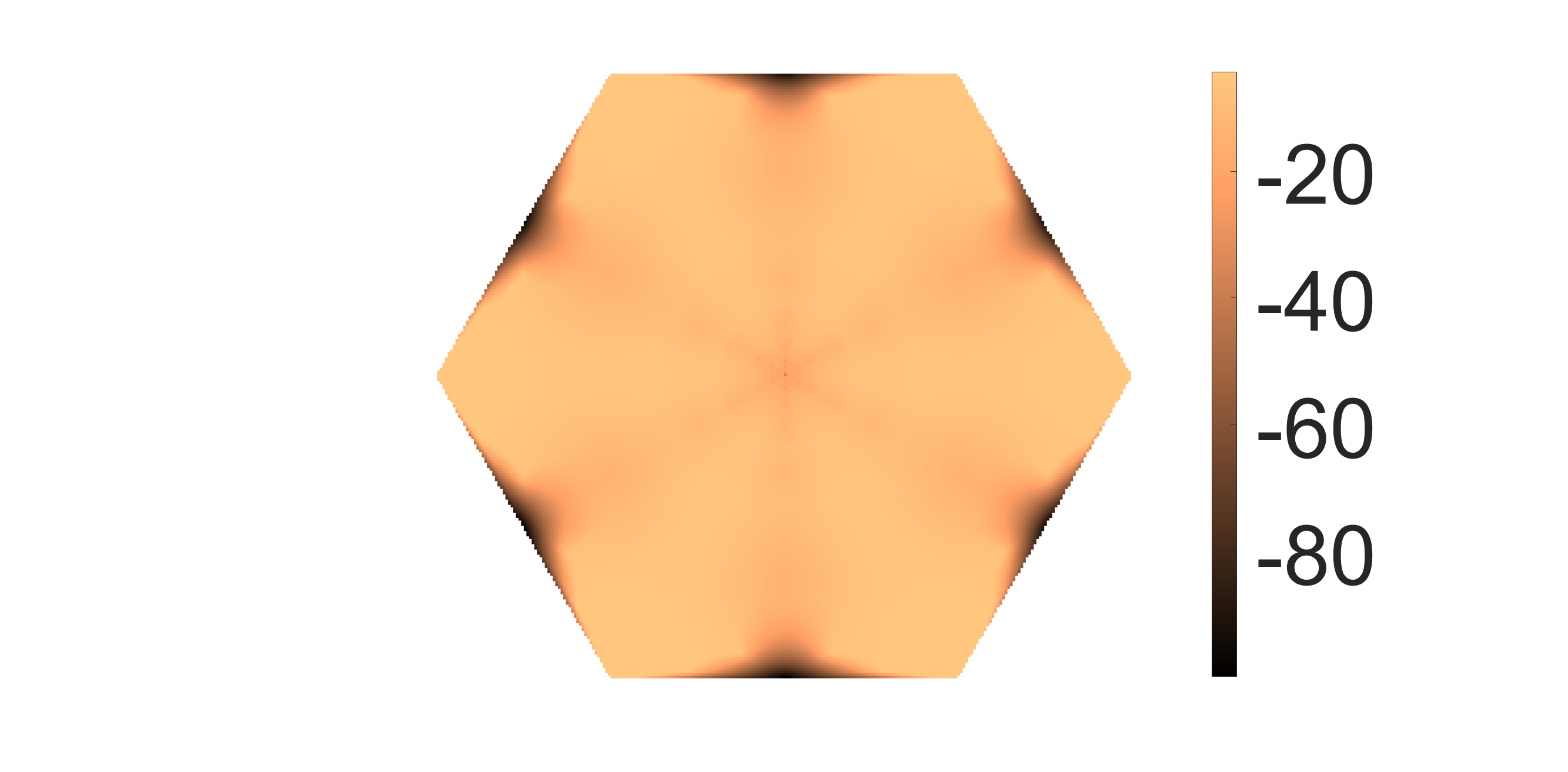}\label{fig:rev5}}};
		\node[above left] at (0.35,2.2) {(d)};
	\end{tikzpicture}
		\begin{tikzpicture}
			\hspace{-23pt}
		\node[anchor=south west,inner sep=0] (image) at (0,0) {\subfigure{\includegraphics[trim=0  200  400  250,clip,width=0.4\textwidth]{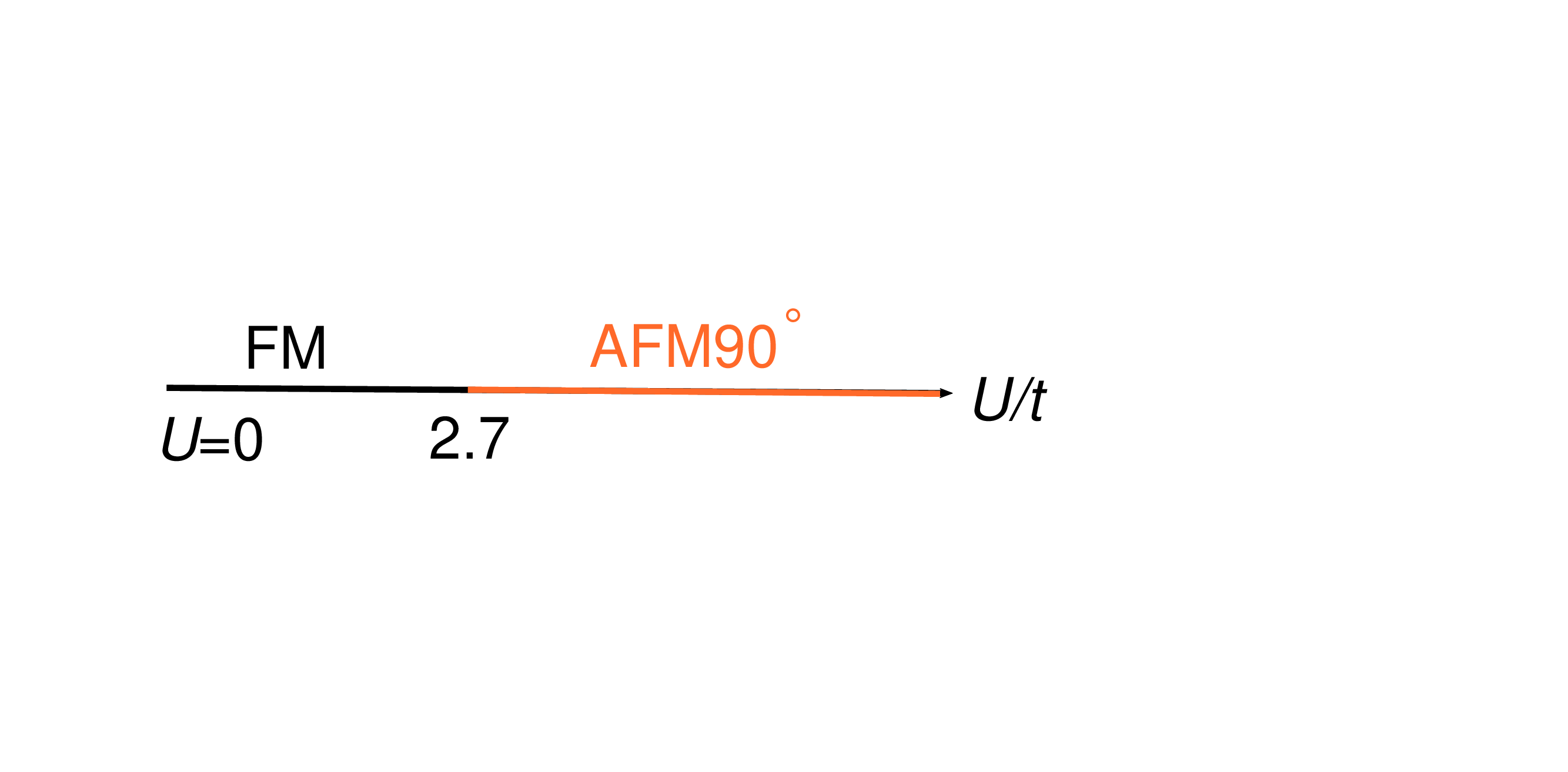}\label{fig:frgphase}}};
		\node[above left] at (0.15,1.2) {(e)};
	\end{tikzpicture}
	\caption{(a) SMFRG flow of leading negative eigenvalues $S$ (plot as $t/S$) at $U=2t$ in the SC (blue), SDW (red), and CDW (yellow) channels, respectively. (b) plots $S(\mathbf{q})$ in the divergent SDW channel. (c,d) are similar to (a,b) but for $U = 5t$. (e) plots the phase diagram from SMFRG calculations.}
	\label{fig:5}
\end{figure}

In Fig.~\ref{fig:flow2}, we plot the inverse of the negative leading eigenvalues $S$ in the three channels, respectively, at a small $U=2t$. We find the SDW channel is the dominant one, much stronger than CDW and SC channels and diverges at first, indicating the magnetic instability. We next examine the effective interaction $V_{\rm SDW}(\mathbf{q})$. The negative leading eivenvalue $S(\mathbf{q})$ is plotted in Fig.~\ref{fig:rev2}, exhibiting the largest peak at $\mathbf{q}=0$, and the corresponding eigenvector is $[1,1,1]/\sqrt{3}$ in the local fermion bilinear basis, indicating a FM instability.
For $U = 5t$, the RG flows are plotted in Fig.~\ref{fig:flow5}. It can be seen the SDW channel still dominates and diverges firstly. But the effective interaction $V_{\rm SDW}$ is different. The leading negative eigenvalue $S(\mathbf{q})$ is shown in Fig.~\ref{fig:rev5}, exhibiting largest peaks at $M$ points. The eigenvectors at the three $M$ points are $[1,0,0]$, $[0,1,0]$, and $[0,0,1]$ (in the local bilineaer basis), respectively, indicating the AFM90$^{\circ}$ instability.

By varying $U$, we find the instability always occurs in the SDW channel and there are only two types of instabilities, FM for small $U$ and AFM90$^{\circ}$ for large $U$, similar to the above RPA but the phase transition point is at around $U_{c1}\approx2.7t$, as shown in Fig.~\ref{fig:frgphase}, larger than the RPA value. This is because FRG takes channel overlap and consequently Coulomb screening effect into account. For even larger $U$ the RG flow diverges too soon at high energy scales. Although this is still suggestive of the development of spin correlations, it violates the RG basis for the truncation of the effective interaction at the level of four-point vertices. Therefore, we resort to VMC simulations in the next section.

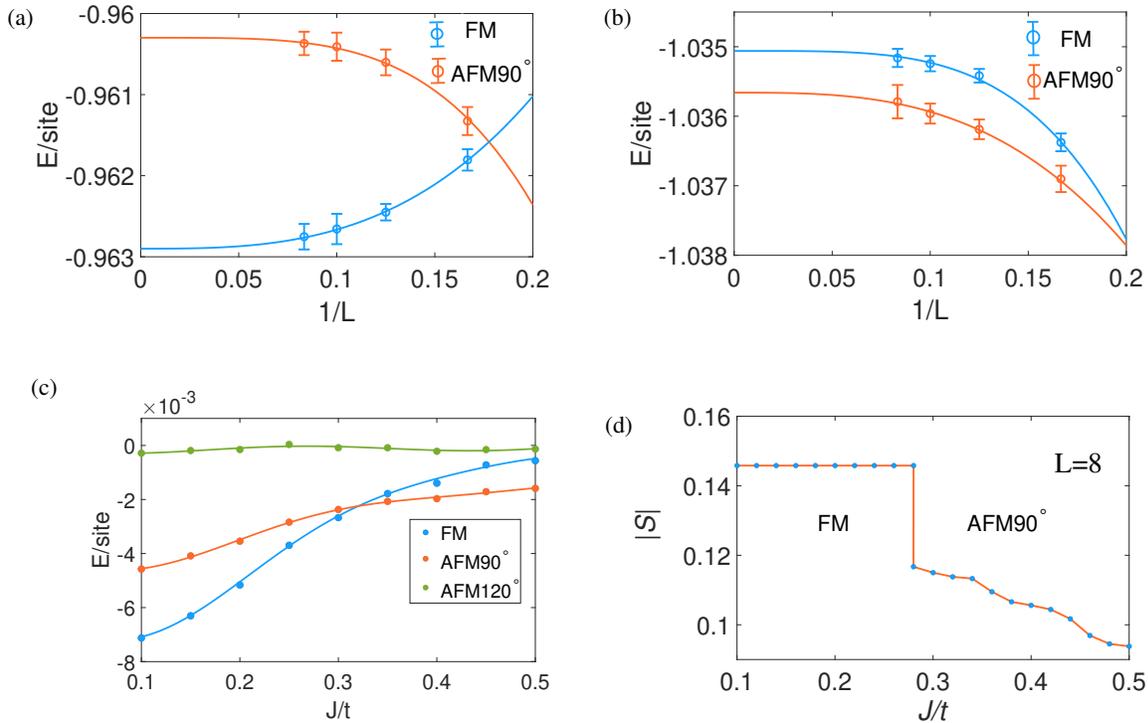
\begin{figure*}
		\begin{tikzpicture}
			\hspace{-20pt}
			\node[anchor=south west,inner sep=0] (image) at (0,0) {\subfigure{\includegraphics[trim=140 50 260 10,clip,width=0.4\textwidth]{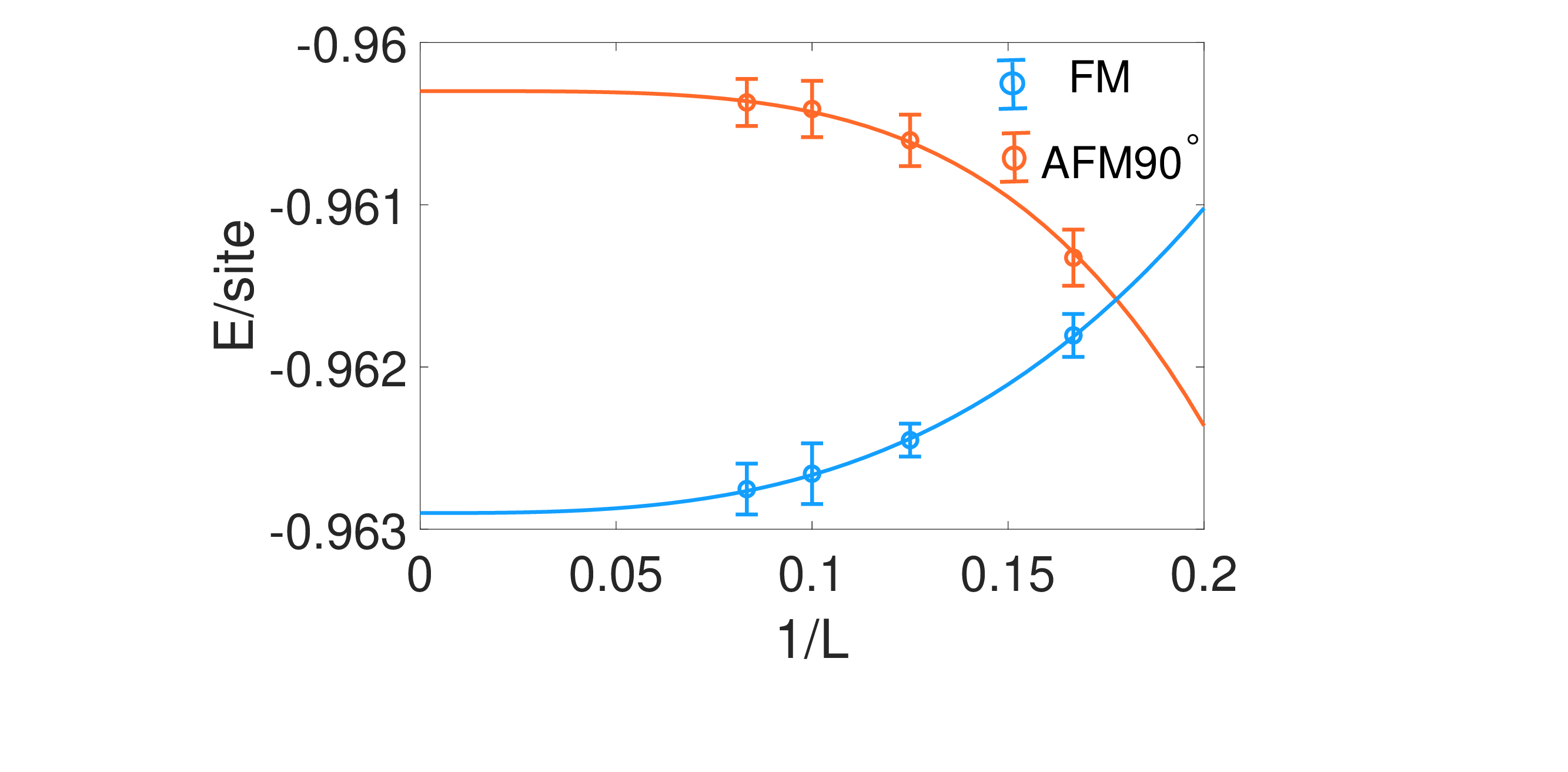}\label{fig:J01}}};
			\node[above left] at (0.35,4.3) {(a)};
		\end{tikzpicture}
		\begin{tikzpicture}
			\hspace{-7pt}

			\node[anchor=south west,inner sep=0] (image) at (0,0) {\subfigure{\includegraphics[trim=150 10 250 40,clip,width=0.4\textwidth]{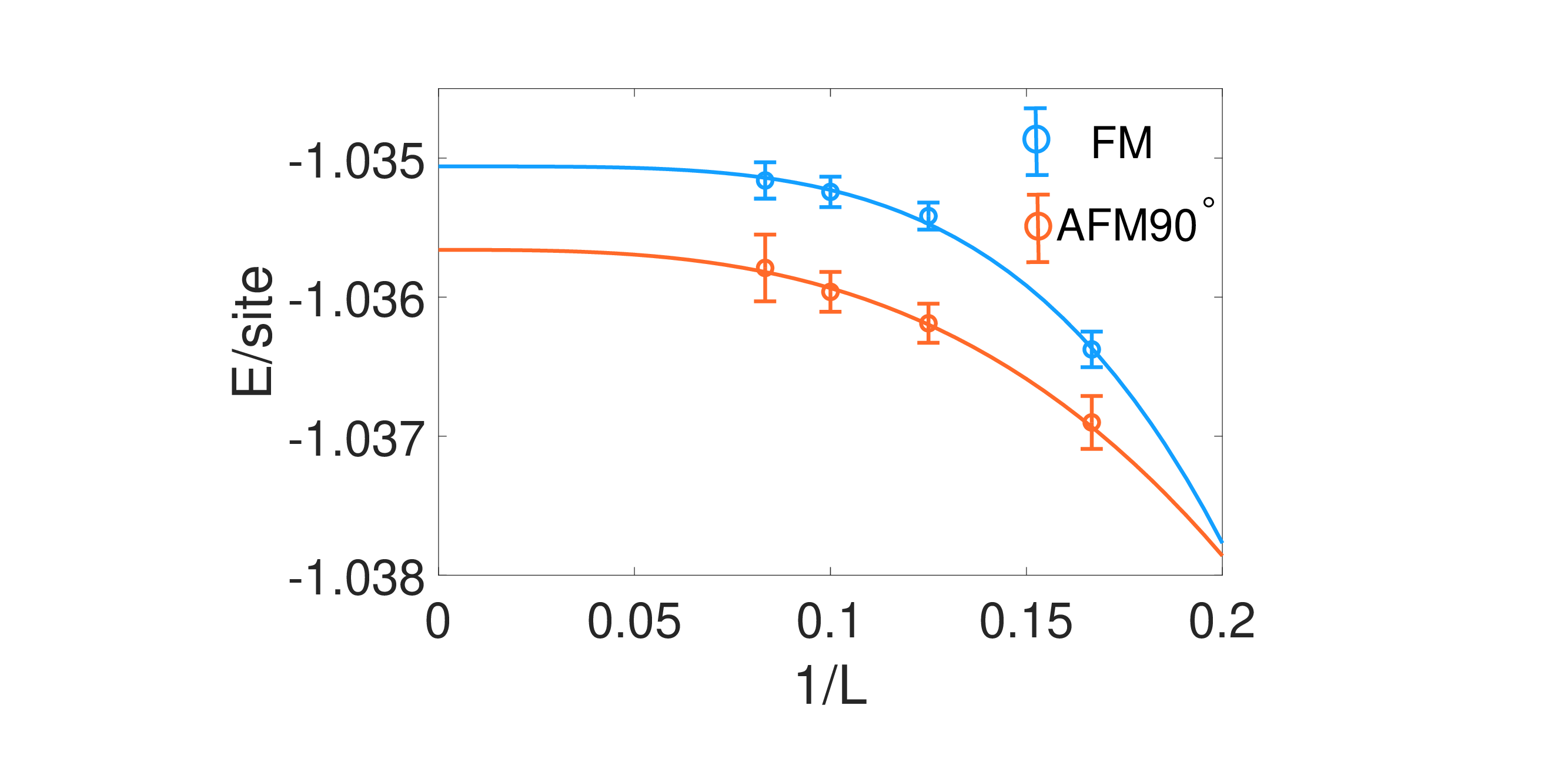}\label{fig:J04}}};
			\node[above left] at (0.45,4.3) {(b)};
		\end{tikzpicture}
		\begin{tikzpicture}
			\hspace{-9pt}
			\node[anchor=south west,inner sep=0] (image) at (0,0) {\subfigure{\includegraphics[trim=0 80 0 90,clip,width=0.365\textwidth]{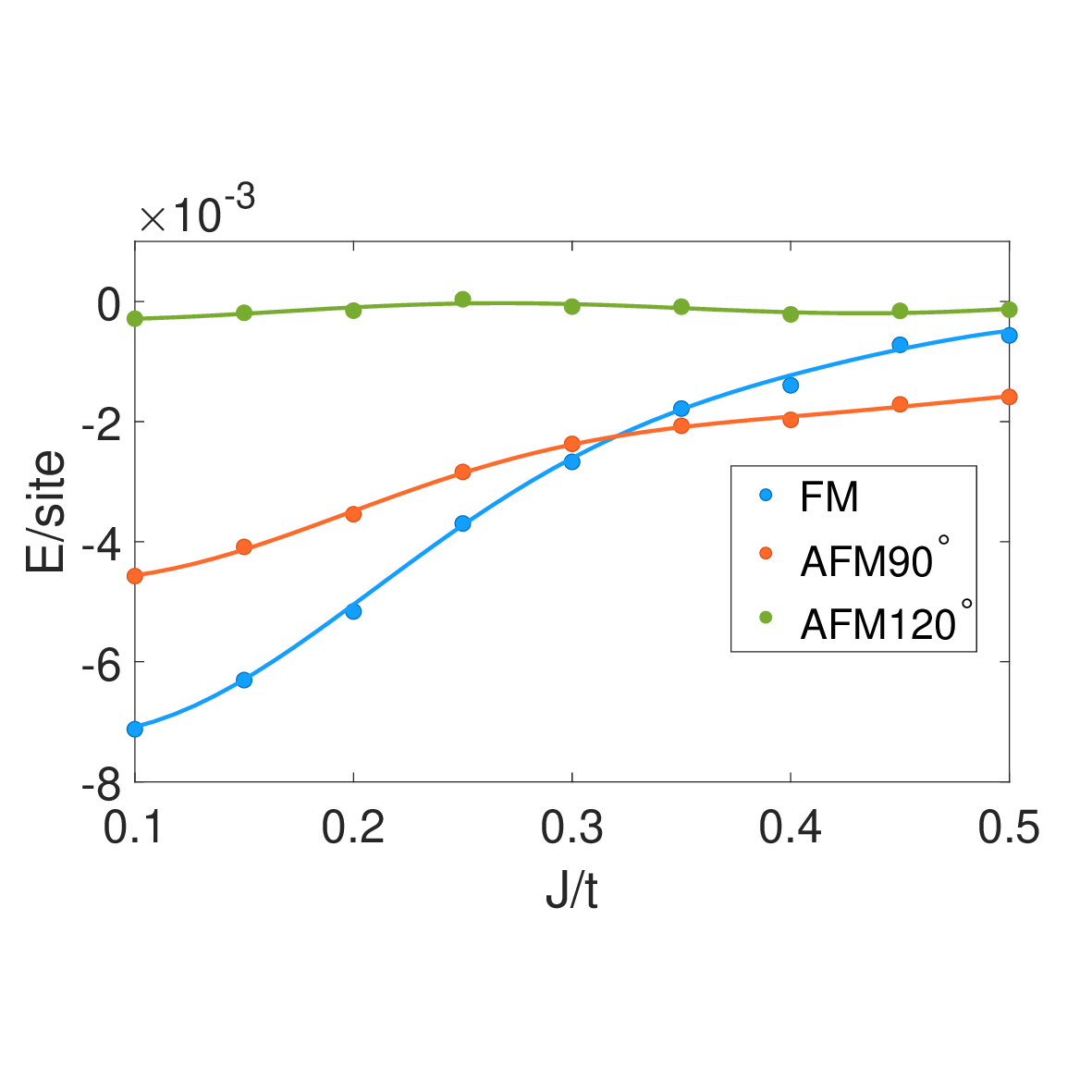}\label{fig:gain}}};
			\node[above left] at (-0.19,4.5) {(c)};
		\end{tikzpicture}
		\begin{tikzpicture}
			\hspace{-2pt}
			\vspace{0pt}
			\node[anchor=south west,inner sep=0] (image) at (0,0) {\subfigure{\includegraphics[trim=110 40 290 10,clip,width=0.4\textwidth]{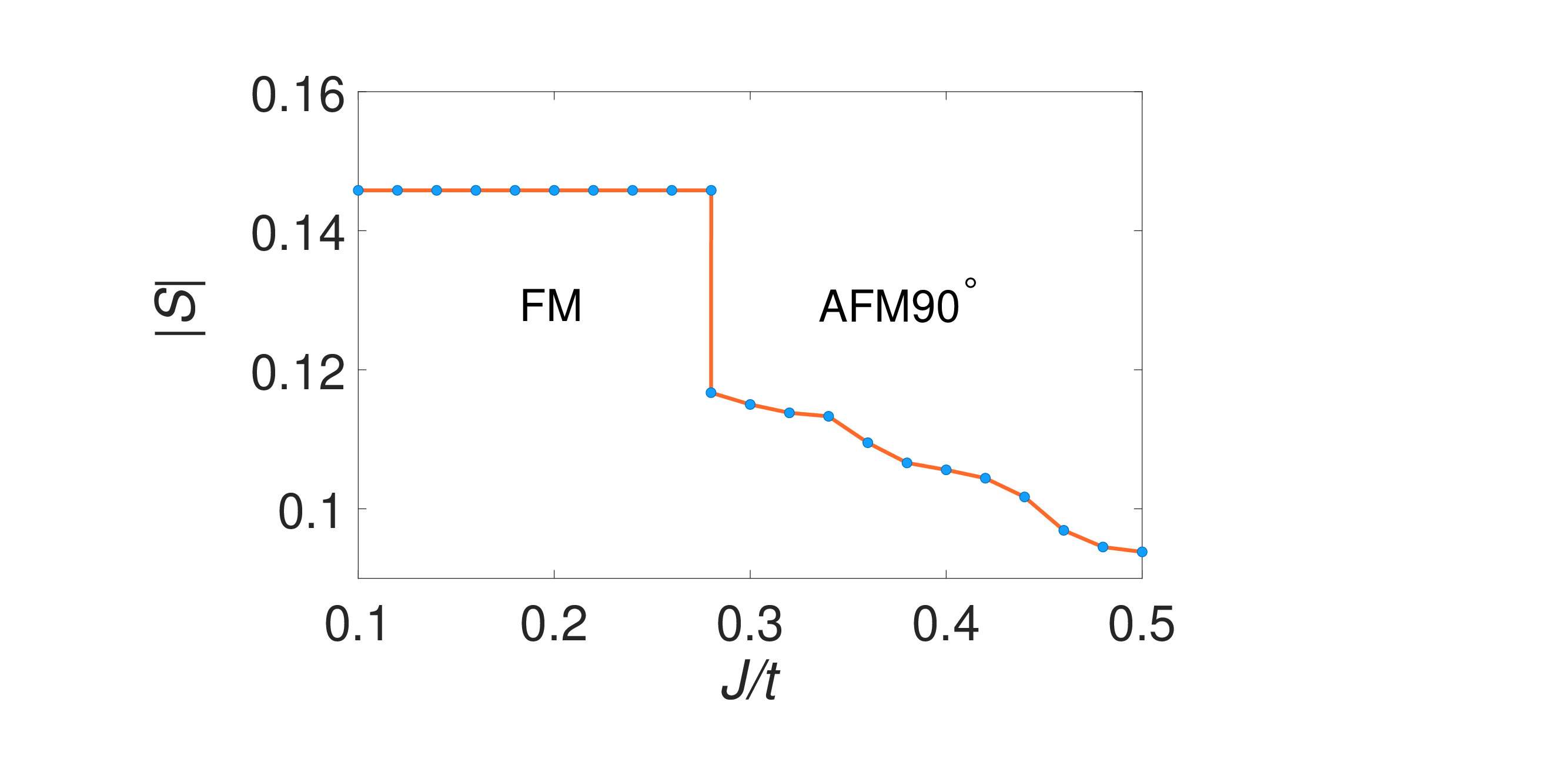}\label{fig:order}}};
            \node[above left,scale=1.2] at (6.5,3.5) {L=8};
			\node[above left] at (0.22,4) {(d)};
		\end{tikzpicture}
\caption{ (a) Optimized VMC energy $E$ (data points) per site in the FM (blue) and AFM90$^\circ$ (red) states versus $1/L$ for $J = 0.1t$. The curves are polynomial fittings to the VMC data. (b) is similar to (a) but for $J = 0.4t$. (c) plots the energy gain $\Delta E$ relative to the normal (paramagnetic) state for the FM (blue), AFM90$^\circ$ (red) and {AFM120$^\circ$ (green) states  in the thermodynamic limit}. The crossing occurs at $J_c\approx 0.32t$, corresponding to $U_{c2}\approx 12.5t$. (d) The ordered spin moment $|\mathbf{S}|$ is plotted as a function of $J$ at $L=8$.}
\label{fig:vmc}
\end{figure*}

\section{\label{VMC}VMC in the Strong coupling limit}

In this section, we consider the case of $U$ larger than the bandwidth, in which it is a {reasonable} choice to replace the Hubbard model with the effective $t$-$J$ model
\begin{equation}
H = -t\sum_{\braket{ij}\sigma}(c^{\dagger}_{i\sigma}c_{j\sigma}+{\rm H.c.}) + J\sum_{\braket{ij}}\left(\mathbf{S}_{i} \cdot\mathbf{S}_{j}-\frac14n_in_j\right),
\end{equation}
where $J = 4t^2/U$, $\mathbf{S}_i$ is the electron spin and $n_i$ is the electron number operator subject to the constrain $n_i\leq 1$.
The chemical potential $\mu$ does not appear since we are going to work in the canonical ensemble. The advantage of the effective model is it captures the spin exchange effect already at the Hamiltonian level. We then
employ VMC to investigate the ground state of the above $t$-$J$ model \cite{vmcbook,vmcreview1}.
As suggested by the previous results at weak to moderate couplings, we choose the Weiss fields $\Delta_{\rm FM}$ for FM and $\Delta_{\rm AFM}$ for AFM90$^{\circ}$ as the variational parameters, which enter the variational Hamiltonian $H_{var}$,
\begin{eqnarray}
H_{var} = H_t - \Delta_{FM} \sum_i S_{iz} -\Delta_{AFM}\sum_i \mathbf{S}_i\cdot\mathbf{e}_i ,
\end{eqnarray}
where $H_t$ is the hopping term, the FM is assumed to be along $z$-direction, and $\mathbf{e}_i$ is the spin orientation at site $i$ for the AFM90$^{\circ}$ configuration.
The variational wave function can be written as $\ket{\psi_{var}} = P_G\ket{\psi_0}$, where $P_G = \Pi_{i}(1-n_{i\uparrow}n_{i\downarrow})$ is the Gutzwiller projection operator to implement no-double occupancy restriction and $\ket{\psi_0}$ is the ground state of $H_{var}$.
For a given $\Delta_{FM}$ or $\Delta_{AFM}$, the physical energy $E = \bra{\psi_{var}}H\ket{\psi_{var}}/\braket{\psi_{var}|\psi_{var}}$ is calculated using the Monte Carlo method.
Then we vary the variational parameter to optimize the physical energy.
In practice, there is an open-shell problem, i.e. the degeneracy of the single particle energies at the Fermi level in $H_{var}$, which prevents us obtaining reasonable $\psi_{var}$. To break the degeneracy, in this work, we add a Peierls phase in the kinetic terms $H_t$ of both the variational and physical Hamiltonians, given by
\begin{equation}
	H_t = -t\sum_{\braket{ij}\sigma}(c^{\dagger}_{i\sigma}c_{j\sigma}e^{i\mathbf{b}\cdot(\mathbf{r}_j-\mathbf{r}_i)}+{\rm H.c.}),
\end{equation}
where $\mathbf{r}_i$ is the coordinate of site $i$, and $\mathbf{b} = (0.37,0.25)\pi / L$ in our simulations, with $L$ the linear size of the system in terms of unitcell. We have checked that different choices of $\mathbf{b}$ do not change the results qualitatively.
We perform VMC simulations on lattices with $L\times L$ unitcells up to $L=12$.

Our VMC results are shown in Fig.~\ref{fig:vmc}.
In (a) and (b), we plot the optimized energy versus $1/L$ for $J=0.1t$ and $0.4t$, respectively. For $J=0.4t$ (or $U=4t^2/J=10t$), the ground state is AFM90$^{\circ}$, consistent with the above FRG result for intermediate $U$. But for $J=0.1t$, corresponding to $U=40t$, the ground state gets back to FM again via a first order transition.
In (c), we plot the energy gain of the two magnetic states relative to the normal (paramagnetic) state {after extrapolation to the thermodynamic limit}, from which we see the ground state changes from FM to AFM90$^{\circ}$ as $J$ increases and the transition point is at $J_c\approx{0.32t}$, corresponding to $U_c\approx{12.5}t$.
{In addition, we have also checked another antiferromagnetic order with the nearby spins differing by 120$^\circ$ \cite{PhysRevB.87.115135}, denoted as AFM120$^\circ$. Its energy is always higher than FM and AFM90$^\circ$, and hence can be safely ruled out as the ground state at the lower van Hove filling.}
In (d), we plot the physical order parameter at $L=8$, defined as the ordered moment per site $|\mathbf{S}|=\sqrt{\braket{S_{ix}}^2+\braket{S_{iy}}^2+\braket{S_{iz}}^2}$. With decreasing $J$, the spin moment $|\mathbf{S}|$ jumps {at $J\approx0.28t$ which is slightly smaller than $J_c$ due to the finite size effect}. This behavior reveals the property of the FM state at small $J$, hence large $U$. The FM state in this limit may be understood as follows: The spin polarization releases the frustration of electron motion from unequal spins, gaining kinetic energy. When this energy gain overwhelms the energy lost from the antiferromagnetic spin exchange, it is energetically favorable to develop FM \cite{PhysRevLett.100.136404}.

\section{\label{sec:level4}summary}
We studied the electronic instabilities in kagome lattice at the lower van Hove filling. We find the FM and the noncollinear AFM90$^{\circ}$ states are the only possible ordered states. The results are summarized as the ground phase diagram in Fig.~\ref{fig:finalphase}. The itinerant FM state is realized at $U<U_{c1}=2.7t$; For $U_{c1}<U<U_{c2}={12.5}t$, the system is in the noncollinear AFM90$^{\circ}$, which is also a Chern insulator supporting quantized anomalous Hall conductance. For $U>U_{c2}$, the FM state revives to gain kinetic energy. The results at the lower van Hove filling are therefore very different to that at the upper one, and can be attributed to the lack of particle-hole symmetry in the band structure with respect to the Dirac point. The results are also different to that in the nearest-neighbor honeycomb lattice, where the particle-hole symmetry is present. Our results enrich the variety of electronic orders in kagome lattices. Experimental realization of the novel AFM90$^{\circ}$ state would be extremely interesting.

\begin{acknowledgments}
This work is supported by National Key R\&D Program of China (Grant No. 2022YFA1403201) and National Natural Science Foundation of China (Grant No. 12374147, No. 12274205, No. 92365203 and No. 11874205).
The numerical calculations were performed at the High Performance Computing Center of Nanjing University.
\end{acknowledgments}

\appendix
\section{\label{MFTdetails} Mean field theory and results}
\begin{figure}
	\begin{tikzpicture}
		\hspace{-9pt}
		\node[anchor=south west,inner sep=0] (image) at (0,0) {\subfigure{\includegraphics[trim=0 40 0 40,clip,width=0.49\textwidth]{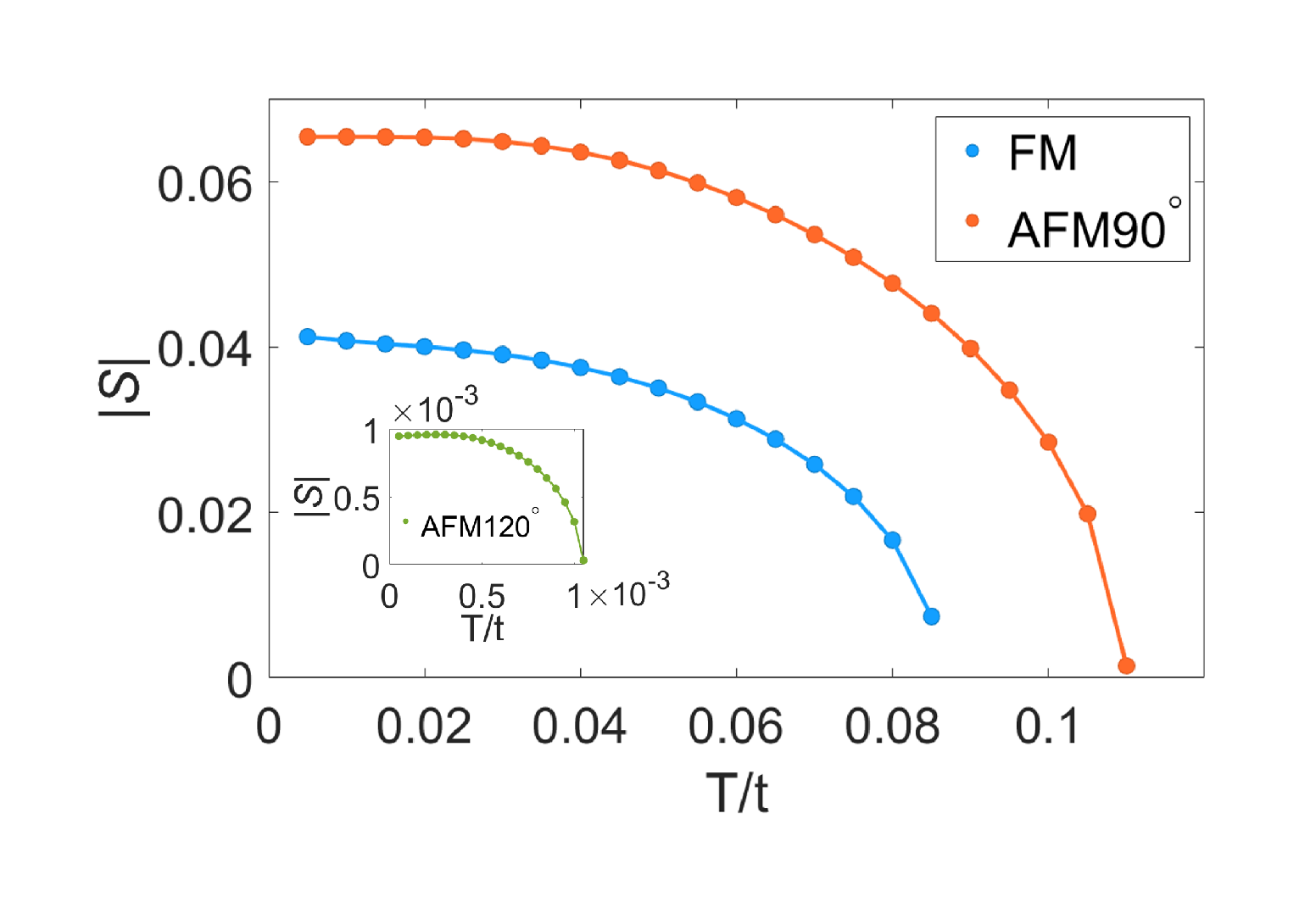}}};
	\end{tikzpicture}
	\caption {The order parameters versus temperature for FM (blue), AFM90$^{\circ}$ (orange), and AFM120$^\circ$ (green) states obtained by the mean field theory at $U=5t$.}
	\label{fig:smmft1}
\end{figure}
First we introduce how we get the mean-field Hamiltonian by variational principle. To get the ground state energy $E_{G}$ of Hamiltonian $H$, we assume a trial wave $\ket{\psi}$ and optimize $E \equiv \braket{\psi|H|\psi}$ through optimizing $\ket{\psi}$. We use $\hat{X_i}$ to represent the Hermitian bilinear operator, and $X_i$ is the corresponding average. Then $E$ is a function of $X = \{X_i\}$ obtained by Wick contraction. With variational principle, we get the stationary equation as
\begin{equation}
	\frac{\delta}{\delta\bra{\psi}}[E - \varepsilon(\braket{\psi|\psi} - 1)] = \sum_i\frac{\partial E}{\partial X_i} \hat{X_i}\ket{\psi} - \varepsilon\ket{\psi} = 0 ,
\end{equation}
where $\varepsilon$ is a Lagrangian multiplier to ensure $\braket{\psi|\psi} = 1$. Then we can rewrite the above equation as
\begin{equation}
	H_{MF}\ket{\psi} = \varepsilon\ket{\psi} ,
\end{equation}
with
\begin{equation}
	H_{MF} = \sum_i \frac{\partial E}{\partial X_i}\hat{X_i} .
\end{equation}
We consider all possible Wick contractions in the Hubbard model and then obtain the mean-field Hamiltonian as
\begin{eqnarray}
	H_{MF} &= &-t\sum_{ij\sigma}(c^\dagger_{i\sigma}c_{j\sigma} +{\rm H.c.}) - 2U\sum_{i}(S_{i}^x\braket{S_{i}^x} \nonumber \\&&+ S_{i}^y\braket{S_{i}^y}
	+S_{i}^z\braket{S_{i}^z}) - \mu N_e + \text{const.},
	 \label{eq:Hmf}
\end{eqnarray}
where $S_{i}^{\alpha} = 1/2c^{\dagger}_{i\sigma}\tau_{\sigma\sigma'}^{\alpha} c_{i\sigma'}(\alpha=x,y,z)$ with $\tau^\alpha$ three Pauli matrices, $\braket{S_{i}^\alpha}$ is the order parameter, and $\text{const.} = U\sum_{i}(\braket{S_i^x}^2+\braket{S_i^y}^2+\braket{S_i^z}^2)$ is a constant. Eq.~\ref{eq:Hmf} gives the mean field decouplings in $S^x$, $S^y$ and $S^z$ channels simultaneously with no ambiguity.

Based on Eq.~\ref{eq:Hmf}, we calculate the order parameters self-consistently. At $U=5t$ with $\chi_0(\mathbf{q})$ diverges at $M$ points, starting from random initial spin configurations, we always obtain the AFM90$^{\circ}$ state as shown in Fig.~\ref{fig:1}. The order parameter $|S|$ is plotted versus temperature in Fig.~\ref{fig:smmft1}.
As comparisons, we also enforce the initial FM configuration to obtain its order parameter, which does vanish at a smaller temperature than AFM90$^{\circ}$ as predicted by FRG.
In addition, we also present the result of AFM120$^\circ$, which is much smaller than FM and thus shown as an enlarged view.


\bibliography{ref}

\begin{thebibliography}{38}%
\makeatletter
\providecommand \@ifxundefined [1]{%
 \@ifx{#1\undefined}
}%
\providecommand \@ifnum [1]{%
 \ifnum #1\expandafter \@firstoftwo
 \else \expandafter \@secondoftwo
 \fi
}%
\providecommand \@ifx [1]{%
 \ifx #1\expandafter \@firstoftwo
 \else \expandafter \@secondoftwo
 \fi
}%
\providecommand \natexlab [1]{#1}%
\providecommand \enquote  [1]{``#1''}%
\providecommand \bibnamefont  [1]{#1}%
\providecommand \bibfnamefont [1]{#1}%
\providecommand \citenamefont [1]{#1}%
\providecommand \href@noop [0]{\@secondoftwo}%
\providecommand \href [0]{\begingroup \@sanitize@url \@href}%
\providecommand \@href[1]{\@@startlink{#1}\@@href}%
\providecommand \@@href[1]{\endgroup#1\@@endlink}%
\providecommand \@sanitize@url [0]{\catcode `\\12\catcode `\$12\catcode
  `\&12\catcode `\#12\catcode `\^12\catcode `\_12\catcode `\%12\relax}%
\providecommand \@@startlink[1]{}%
\providecommand \@@endlink[0]{}%
\providecommand \url  [0]{\begingroup\@sanitize@url \@url }%
\providecommand \@url [1]{\endgroup\@href {#1}{\urlprefix }}%
\providecommand \urlprefix  [0]{URL }%
\providecommand \Eprint [0]{\href }%
\providecommand \doibase [0]{https://doi.org/}%
\providecommand \selectlanguage [0]{\@gobble}%
\providecommand \bibinfo  [0]{\@secondoftwo}%
\providecommand \bibfield  [0]{\@secondoftwo}%
\providecommand \translation [1]{[#1]}%
\providecommand \BibitemOpen [0]{}%
\providecommand \bibitemStop [0]{}%
\providecommand \bibitemNoStop [0]{.\EOS\space}%
\providecommand \EOS [0]{\spacefactor3000\relax}%
\providecommand \BibitemShut  [1]{\csname bibitem#1\endcsname}%
\let\auto@bib@innerbib\@empty
\bibitem [{\citenamefont {Yin}\ \emph {et~al.}(2022)\citenamefont {Yin},
  \citenamefont {Lian},\ and\ \citenamefont {Hasan}}]{Yin_N_2022}%
  \BibitemOpen
  \bibfield  {author} {\bibinfo {author} {\bibfnamefont {J.-X.}\ \bibnamefont
  {Yin}}, \bibinfo {author} {\bibfnamefont {B.}~\bibnamefont {Lian}},\ and\
  \bibinfo {author} {\bibfnamefont {M.~Z.}\ \bibnamefont {Hasan}},\ }\bibfield
  {title} {\bibinfo {title} {Topological kagome magnets and superconductors},\
  }\href {https://doi.org/10.1038/s41586-022-05516-0} {\bibfield  {journal}
  {\bibinfo  {journal} {Nature}\ }\textbf {\bibinfo {volume} {612}},\ \bibinfo
  {pages} {647} (\bibinfo {year} {2022})}\BibitemShut {NoStop}%
\bibitem [{\citenamefont {Liu}\ \emph {et~al.}(2020)\citenamefont {Liu},
  \citenamefont {Li}, \citenamefont {Wang}, \citenamefont {Wang}, \citenamefont
  {Wen}, \citenamefont {Jiang}, \citenamefont {Lu}, \citenamefont {Yan},
  \citenamefont {Huang}, \citenamefont {Shen}, \citenamefont {Yin},
  \citenamefont {Wang}, \citenamefont {Yin}, \citenamefont {Lei},\ and\
  \citenamefont {Wang}}]{Liu2020}%
  \BibitemOpen
  \bibfield  {author} {\bibinfo {author} {\bibfnamefont {Z.}~\bibnamefont
  {Liu}}, \bibinfo {author} {\bibfnamefont {M.}~\bibnamefont {Li}}, \bibinfo
  {author} {\bibfnamefont {Q.}~\bibnamefont {Wang}}, \bibinfo {author}
  {\bibfnamefont {G.}~\bibnamefont {Wang}}, \bibinfo {author} {\bibfnamefont
  {C.}~\bibnamefont {Wen}}, \bibinfo {author} {\bibfnamefont {K.}~\bibnamefont
  {Jiang}}, \bibinfo {author} {\bibfnamefont {X.}~\bibnamefont {Lu}}, \bibinfo
  {author} {\bibfnamefont {S.}~\bibnamefont {Yan}}, \bibinfo {author}
  {\bibfnamefont {Y.}~\bibnamefont {Huang}}, \bibinfo {author} {\bibfnamefont
  {D.}~\bibnamefont {Shen}}, \bibinfo {author} {\bibfnamefont {J.-X.}\
  \bibnamefont {Yin}}, \bibinfo {author} {\bibfnamefont {Z.}~\bibnamefont
  {Wang}}, \bibinfo {author} {\bibfnamefont {Z.}~\bibnamefont {Yin}}, \bibinfo
  {author} {\bibfnamefont {H.}~\bibnamefont {Lei}},\ and\ \bibinfo {author}
  {\bibfnamefont {S.}~\bibnamefont {Wang}},\ }\bibfield  {title} {\bibinfo
  {title} {Orbital-selective dirac fermions and extremely flat bands in
  frustrated kagome-lattice metal cosn},\ }\href
  {https://doi.org/10.1038/s41467-020-17462-4} {\bibfield  {journal} {\bibinfo
  {journal} {Nature Communications}\ }\textbf {\bibinfo {volume} {11}},\
  \bibinfo {pages} {4002} (\bibinfo {year} {2020})}\BibitemShut {NoStop}%
\bibitem [{\citenamefont {Zhang}\ \emph {et~al.}(2022)\citenamefont {Zhang},
  \citenamefont {Koo}, \citenamefont {Xu}, \citenamefont {Sretenovic},
  \citenamefont {Yan},\ and\ \citenamefont {Ke}}]{Zhang2022}%
  \BibitemOpen
  \bibfield  {author} {\bibinfo {author} {\bibfnamefont {H.}~\bibnamefont
  {Zhang}}, \bibinfo {author} {\bibfnamefont {J.}~\bibnamefont {Koo}}, \bibinfo
  {author} {\bibfnamefont {C.}~\bibnamefont {Xu}}, \bibinfo {author}
  {\bibfnamefont {M.}~\bibnamefont {Sretenovic}}, \bibinfo {author}
  {\bibfnamefont {B.}~\bibnamefont {Yan}},\ and\ \bibinfo {author}
  {\bibfnamefont {X.}~\bibnamefont {Ke}},\ }\bibfield  {title} {\bibinfo
  {title} {Exchange-biased topological transverse thermoelectric effects in a
  kagome ferrimagnet},\ }\href {https://doi.org/10.1038/s41467-022-28733-7}
  {\bibfield  {journal} {\bibinfo  {journal} {Nature Communications}\ }\textbf
  {\bibinfo {volume} {13}},\ \bibinfo {pages} {1091} (\bibinfo {year}
  {2022})}\BibitemShut {NoStop}%
\bibitem [{\citenamefont {Yin}\ \emph {et~al.}(2020)\citenamefont {Yin},
  \citenamefont {Shumiya}, \citenamefont {Mardanya}, \citenamefont {Wang},
  \citenamefont {Zhang}, \citenamefont {Tien}, \citenamefont {Multer},
  \citenamefont {Jiang}, \citenamefont {Cheng}, \citenamefont {Yao},
  \citenamefont {Wu}, \citenamefont {Wu}, \citenamefont {Deng}, \citenamefont
  {Ye}, \citenamefont {He}, \citenamefont {Chang}, \citenamefont {Liu},
  \citenamefont {Jiang}, \citenamefont {Wang}, \citenamefont {Neupert},
  \citenamefont {Agarwal}, \citenamefont {Chang}, \citenamefont {Chu},
  \citenamefont {Lei},\ and\ \citenamefont {Hasan}}]{Yin2020}%
  \BibitemOpen
  \bibfield  {author} {\bibinfo {author} {\bibfnamefont {J.-X.}\ \bibnamefont
  {Yin}}, \bibinfo {author} {\bibfnamefont {N.}~\bibnamefont {Shumiya}},
  \bibinfo {author} {\bibfnamefont {S.}~\bibnamefont {Mardanya}}, \bibinfo
  {author} {\bibfnamefont {Q.}~\bibnamefont {Wang}}, \bibinfo {author}
  {\bibfnamefont {S.~S.}\ \bibnamefont {Zhang}}, \bibinfo {author}
  {\bibfnamefont {H.-J.}\ \bibnamefont {Tien}}, \bibinfo {author}
  {\bibfnamefont {D.}~\bibnamefont {Multer}}, \bibinfo {author} {\bibfnamefont
  {Y.}~\bibnamefont {Jiang}}, \bibinfo {author} {\bibfnamefont
  {G.}~\bibnamefont {Cheng}}, \bibinfo {author} {\bibfnamefont
  {N.}~\bibnamefont {Yao}}, \bibinfo {author} {\bibfnamefont {S.}~\bibnamefont
  {Wu}}, \bibinfo {author} {\bibfnamefont {D.}~\bibnamefont {Wu}}, \bibinfo
  {author} {\bibfnamefont {L.}~\bibnamefont {Deng}}, \bibinfo {author}
  {\bibfnamefont {Z.}~\bibnamefont {Ye}}, \bibinfo {author} {\bibfnamefont
  {R.}~\bibnamefont {He}}, \bibinfo {author} {\bibfnamefont {G.}~\bibnamefont
  {Chang}}, \bibinfo {author} {\bibfnamefont {Z.}~\bibnamefont {Liu}}, \bibinfo
  {author} {\bibfnamefont {K.}~\bibnamefont {Jiang}}, \bibinfo {author}
  {\bibfnamefont {Z.}~\bibnamefont {Wang}}, \bibinfo {author} {\bibfnamefont
  {T.}~\bibnamefont {Neupert}}, \bibinfo {author} {\bibfnamefont
  {A.}~\bibnamefont {Agarwal}}, \bibinfo {author} {\bibfnamefont {T.-R.}\
  \bibnamefont {Chang}}, \bibinfo {author} {\bibfnamefont {C.-W.}\ \bibnamefont
  {Chu}}, \bibinfo {author} {\bibfnamefont {H.}~\bibnamefont {Lei}},\ and\
  \bibinfo {author} {\bibfnamefont {M.~Z.}\ \bibnamefont {Hasan}},\ }\bibfield
  {title} {\bibinfo {title} {Fermion--boson many-body interplay in a frustrated
  kagome paramagnet},\ }\href {https://doi.org/10.1038/s41467-020-17464-2}
  {\bibfield  {journal} {\bibinfo  {journal} {Nature Communications}\ }\textbf
  {\bibinfo {volume} {11}},\ \bibinfo {pages} {4003} (\bibinfo {year}
  {2020})}\BibitemShut {NoStop}%
\bibitem [{\citenamefont {Wakao}\ \emph {et~al.}(2020)\citenamefont {Wakao},
  \citenamefont {Yoshida}, \citenamefont {Araki}, \citenamefont {Mizoguchi},\
  and\ \citenamefont {Hatsugai}}]{PhysRevB.101.094107}%
  \BibitemOpen
  \bibfield  {author} {\bibinfo {author} {\bibfnamefont {H.}~\bibnamefont
  {Wakao}}, \bibinfo {author} {\bibfnamefont {T.}~\bibnamefont {Yoshida}},
  \bibinfo {author} {\bibfnamefont {H.}~\bibnamefont {Araki}}, \bibinfo
  {author} {\bibfnamefont {T.}~\bibnamefont {Mizoguchi}},\ and\ \bibinfo
  {author} {\bibfnamefont {Y.}~\bibnamefont {Hatsugai}},\ }\bibfield  {title}
  {\bibinfo {title} {Higher-order topological phases in a spring-mass model on
  a breathing kagome lattice},\ }\href
  {https://doi.org/10.1103/PhysRevB.101.094107} {\bibfield  {journal} {\bibinfo
   {journal} {Phys. Rev. B}\ }\textbf {\bibinfo {volume} {101}},\ \bibinfo
  {pages} {094107} (\bibinfo {year} {2020})}\BibitemShut {NoStop}%
\bibitem [{\citenamefont {Lopez-Bezanilla}\ \emph {et~al.}(2023)\citenamefont
  {Lopez-Bezanilla}, \citenamefont {Raymond}, \citenamefont {Boothby},
  \citenamefont {Carrasquilla}, \citenamefont {Nisoli},\ and\ \citenamefont
  {King}}]{Lopez-Bezanilla2023}%
  \BibitemOpen
  \bibfield  {author} {\bibinfo {author} {\bibfnamefont {A.}~\bibnamefont
  {Lopez-Bezanilla}}, \bibinfo {author} {\bibfnamefont {J.}~\bibnamefont
  {Raymond}}, \bibinfo {author} {\bibfnamefont {K.}~\bibnamefont {Boothby}},
  \bibinfo {author} {\bibfnamefont {J.}~\bibnamefont {Carrasquilla}}, \bibinfo
  {author} {\bibfnamefont {C.}~\bibnamefont {Nisoli}},\ and\ \bibinfo {author}
  {\bibfnamefont {A.~D.}\ \bibnamefont {King}},\ }\bibfield  {title} {\bibinfo
  {title} {Kagome qubit ice},\ }\href
  {https://doi.org/10.1038/s41467-023-36760-1} {\bibfield  {journal} {\bibinfo
  {journal} {Nature Communications}\ }\textbf {\bibinfo {volume} {14}},\
  \bibinfo {pages} {1105} (\bibinfo {year} {2023})}\BibitemShut {NoStop}%
\bibitem [{\citenamefont {Jiang}\ \emph {et~al.}(2008)\citenamefont {Jiang},
  \citenamefont {Weng},\ and\ \citenamefont {Sheng}}]{PhysRevLett.101.117203}%
  \BibitemOpen
  \bibfield  {author} {\bibinfo {author} {\bibfnamefont {H.~C.}\ \bibnamefont
  {Jiang}}, \bibinfo {author} {\bibfnamefont {Z.~Y.}\ \bibnamefont {Weng}},\
  and\ \bibinfo {author} {\bibfnamefont {D.~N.}\ \bibnamefont {Sheng}},\
  }\bibfield  {title} {\bibinfo {title} {Density matrix renormalization group
  numerical study of the kagome antiferromagnet},\ }\href
  {https://doi.org/10.1103/PhysRevLett.101.117203} {\bibfield  {journal}
  {\bibinfo  {journal} {Phys. Rev. Lett.}\ }\textbf {\bibinfo {volume} {101}},\
  \bibinfo {pages} {117203} (\bibinfo {year} {2008})}\BibitemShut {NoStop}%
\bibitem [{\citenamefont {Singh}\ and\ \citenamefont
  {Huse}(2007)}]{PhysRevB.76.180407}%
  \BibitemOpen
  \bibfield  {author} {\bibinfo {author} {\bibfnamefont {R.~R.~P.}\
  \bibnamefont {Singh}}\ and\ \bibinfo {author} {\bibfnamefont {D.~A.}\
  \bibnamefont {Huse}},\ }\bibfield  {title} {\bibinfo {title} {Ground state of
  the spin-1/2 kagome-lattice heisenberg antiferromagnet},\ }\href
  {https://doi.org/10.1103/PhysRevB.76.180407} {\bibfield  {journal} {\bibinfo
  {journal} {Phys. Rev. B}\ }\textbf {\bibinfo {volume} {76}},\ \bibinfo
  {pages} {180407} (\bibinfo {year} {2007})}\BibitemShut {NoStop}%
\bibitem [{\citenamefont {Ran}\ \emph {et~al.}(2007)\citenamefont {Ran},
  \citenamefont {Hermele}, \citenamefont {Lee},\ and\ \citenamefont
  {Wen}}]{PhysRevLett.98.117205}%
  \BibitemOpen
  \bibfield  {author} {\bibinfo {author} {\bibfnamefont {Y.}~\bibnamefont
  {Ran}}, \bibinfo {author} {\bibfnamefont {M.}~\bibnamefont {Hermele}},
  \bibinfo {author} {\bibfnamefont {P.~A.}\ \bibnamefont {Lee}},\ and\ \bibinfo
  {author} {\bibfnamefont {X.-G.}\ \bibnamefont {Wen}},\ }\bibfield  {title}
  {\bibinfo {title} {Projected-wave-function study of the spin-$1/2$ heisenberg
  model on the kagom\'e lattice},\ }\href
  {https://doi.org/10.1103/PhysRevLett.98.117205} {\bibfield  {journal}
  {\bibinfo  {journal} {Phys. Rev. Lett.}\ }\textbf {\bibinfo {volume} {98}},\
  \bibinfo {pages} {117205} (\bibinfo {year} {2007})}\BibitemShut {NoStop}%
\bibitem [{\citenamefont {Yan}\ \emph {et~al.}(2011)\citenamefont {Yan},
  \citenamefont {Huse},\ and\ \citenamefont {White}}]{science}%
  \BibitemOpen
  \bibfield  {author} {\bibinfo {author} {\bibfnamefont {S.}~\bibnamefont
  {Yan}}, \bibinfo {author} {\bibfnamefont {D.~A.}\ \bibnamefont {Huse}},\ and\
  \bibinfo {author} {\bibfnamefont {S.~R.}\ \bibnamefont {White}},\ }\bibfield
  {title} {\bibinfo {title} {Spin-liquid ground state of the {S= 1/2} kagome
  heisenberg antiferromagnet},\ }\href
  {https://doi.org/10.1126/science.1201080} {\bibfield  {journal} {\bibinfo
  {journal} {Science}\ }\textbf {\bibinfo {volume} {332}},\ \bibinfo {pages}
  {1173} (\bibinfo {year} {2011})}\BibitemShut {NoStop}%
\bibitem [{\citenamefont {Liao}\ \emph {et~al.}(2017)\citenamefont {Liao},
  \citenamefont {Xie}, \citenamefont {Chen}, \citenamefont {Liu}, \citenamefont
  {Xie}, \citenamefont {Huang}, \citenamefont {Normand},\ and\ \citenamefont
  {Xiang}}]{PhysRevLett.118.137202}%
  \BibitemOpen
  \bibfield  {author} {\bibinfo {author} {\bibfnamefont {H.~J.}\ \bibnamefont
  {Liao}}, \bibinfo {author} {\bibfnamefont {Z.~Y.}\ \bibnamefont {Xie}},
  \bibinfo {author} {\bibfnamefont {J.}~\bibnamefont {Chen}}, \bibinfo {author}
  {\bibfnamefont {Z.~Y.}\ \bibnamefont {Liu}}, \bibinfo {author} {\bibfnamefont
  {H.~D.}\ \bibnamefont {Xie}}, \bibinfo {author} {\bibfnamefont {R.~Z.}\
  \bibnamefont {Huang}}, \bibinfo {author} {\bibfnamefont {B.}~\bibnamefont
  {Normand}},\ and\ \bibinfo {author} {\bibfnamefont {T.}~\bibnamefont
  {Xiang}},\ }\bibfield  {title} {\bibinfo {title} {Gapless spin-liquid ground
  state in the {$S=1/2$} kagome antiferromagnet},\ }\href
  {https://doi.org/10.1103/PhysRevLett.118.137202} {\bibfield  {journal}
  {\bibinfo  {journal} {Phys. Rev. Lett.}\ }\textbf {\bibinfo {volume} {118}},\
  \bibinfo {pages} {137202} (\bibinfo {year} {2017})}\BibitemShut {NoStop}%
\bibitem [{\citenamefont {Wietek}\ \emph {et~al.}(2015)\citenamefont {Wietek},
  \citenamefont {Sterdyniak},\ and\ \citenamefont
  {L\"auchli}}]{PhysRevB.92.125122}%
  \BibitemOpen
  \bibfield  {author} {\bibinfo {author} {\bibfnamefont {A.}~\bibnamefont
  {Wietek}}, \bibinfo {author} {\bibfnamefont {A.}~\bibnamefont {Sterdyniak}},\
  and\ \bibinfo {author} {\bibfnamefont {A.~M.}\ \bibnamefont {L\"auchli}},\
  }\bibfield  {title} {\bibinfo {title} {Nature of chiral spin liquids on the
  kagome lattice},\ }\href {https://doi.org/10.1103/PhysRevB.92.125122}
  {\bibfield  {journal} {\bibinfo  {journal} {Phys. Rev. B}\ }\textbf {\bibinfo
  {volume} {92}},\ \bibinfo {pages} {125122} (\bibinfo {year}
  {2015})}\BibitemShut {NoStop}%
\bibitem [{\citenamefont {Hu}\ \emph {et~al.}(2015)\citenamefont {Hu},
  \citenamefont {Zhu}, \citenamefont {Zhang}, \citenamefont {Gong},
  \citenamefont {Becca},\ and\ \citenamefont {Sheng}}]{PhysRevB.91.041124}%
  \BibitemOpen
  \bibfield  {author} {\bibinfo {author} {\bibfnamefont {W.-J.}\ \bibnamefont
  {Hu}}, \bibinfo {author} {\bibfnamefont {W.}~\bibnamefont {Zhu}}, \bibinfo
  {author} {\bibfnamefont {Y.}~\bibnamefont {Zhang}}, \bibinfo {author}
  {\bibfnamefont {S.}~\bibnamefont {Gong}}, \bibinfo {author} {\bibfnamefont
  {F.}~\bibnamefont {Becca}},\ and\ \bibinfo {author} {\bibfnamefont {D.~N.}\
  \bibnamefont {Sheng}},\ }\bibfield  {title} {\bibinfo {title} {Variational
  monte carlo study of a chiral spin liquid in the extended heisenberg model on
  the kagome lattice},\ }\href {https://doi.org/10.1103/PhysRevB.91.041124}
  {\bibfield  {journal} {\bibinfo  {journal} {Phys. Rev. B}\ }\textbf {\bibinfo
  {volume} {91}},\ \bibinfo {pages} {041124} (\bibinfo {year}
  {2015})}\BibitemShut {NoStop}%
\bibitem [{\citenamefont {Zhou}\ \emph {et~al.}(2017)\citenamefont {Zhou},
  \citenamefont {Kanoda},\ and\ \citenamefont {Ng}}]{Zhou_RMP_2017}%
  \BibitemOpen
  \bibfield  {author} {\bibinfo {author} {\bibfnamefont {Y.}~\bibnamefont
  {Zhou}}, \bibinfo {author} {\bibfnamefont {K.}~\bibnamefont {Kanoda}},\ and\
  \bibinfo {author} {\bibfnamefont {T.-K.}\ \bibnamefont {Ng}},\ }\bibfield
  {title} {\bibinfo {title} {Quantum spin liquid states},\ }\href
  {https://doi.org/10.1103/RevModPhys.89.025003} {\bibfield  {journal}
  {\bibinfo  {journal} {Rev. Mod. Phys.}\ }\textbf {\bibinfo {volume} {89}},\
  \bibinfo {pages} {025003} (\bibinfo {year} {2017})}\BibitemShut {NoStop}%
\bibitem [{\citenamefont {Bosse}\ and\ \citenamefont
  {Montanaro}(2022)}]{PhysRevB.105.094409}%
  \BibitemOpen
  \bibfield  {author} {\bibinfo {author} {\bibfnamefont {J.~L.}\ \bibnamefont
  {Bosse}}\ and\ \bibinfo {author} {\bibfnamefont {A.}~\bibnamefont
  {Montanaro}},\ }\bibfield  {title} {\bibinfo {title} {Probing ground-state
  properties of the kagome antiferromagnetic heisenberg model using the
  variational quantum eigensolver},\ }\href
  {https://doi.org/10.1103/PhysRevB.105.094409} {\bibfield  {journal} {\bibinfo
   {journal} {Phys. Rev. B}\ }\textbf {\bibinfo {volume} {105}},\ \bibinfo
  {pages} {094409} (\bibinfo {year} {2022})}\BibitemShut {NoStop}%
\bibitem [{\citenamefont {Kiese}\ \emph {et~al.}(2023)\citenamefont {Kiese},
  \citenamefont {Ferrari}, \citenamefont {Astrakhantsev}, \citenamefont
  {Niggemann}, \citenamefont {Ghosh}, \citenamefont {M\"uller}, \citenamefont
  {Thomale}, \citenamefont {Neupert}, \citenamefont {Reuther}, \citenamefont
  {Gingras}, \citenamefont {Trebst},\ and\ \citenamefont
  {Iqbal}}]{PhysRevResearch.5.L012025}%
  \BibitemOpen
  \bibfield  {author} {\bibinfo {author} {\bibfnamefont {D.}~\bibnamefont
  {Kiese}}, \bibinfo {author} {\bibfnamefont {F.}~\bibnamefont {Ferrari}},
  \bibinfo {author} {\bibfnamefont {N.}~\bibnamefont {Astrakhantsev}}, \bibinfo
  {author} {\bibfnamefont {N.}~\bibnamefont {Niggemann}}, \bibinfo {author}
  {\bibfnamefont {P.}~\bibnamefont {Ghosh}}, \bibinfo {author} {\bibfnamefont
  {T.}~\bibnamefont {M\"uller}}, \bibinfo {author} {\bibfnamefont
  {R.}~\bibnamefont {Thomale}}, \bibinfo {author} {\bibfnamefont
  {T.}~\bibnamefont {Neupert}}, \bibinfo {author} {\bibfnamefont
  {J.}~\bibnamefont {Reuther}}, \bibinfo {author} {\bibfnamefont {M.~J.~P.}\
  \bibnamefont {Gingras}}, \bibinfo {author} {\bibfnamefont {S.}~\bibnamefont
  {Trebst}},\ and\ \bibinfo {author} {\bibfnamefont {Y.}~\bibnamefont
  {Iqbal}},\ }\bibfield  {title} {\bibinfo {title} {Pinch-points to half-moons
  and up in the stars: The kagome skymap},\ }\href
  {https://doi.org/10.1103/PhysRevResearch.5.L012025} {\bibfield  {journal}
  {\bibinfo  {journal} {Phys. Rev. Research}\ }\textbf {\bibinfo {volume}
  {5}},\ \bibinfo {pages} {L012025} (\bibinfo {year} {2023})}\BibitemShut
  {NoStop}%
\bibitem [{\citenamefont {Iqbal}\ \emph {et~al.}(2013)\citenamefont {Iqbal},
  \citenamefont {Becca}, \citenamefont {Sorella},\ and\ \citenamefont
  {Poilblanc}}]{prb065405}%
  \BibitemOpen
  \bibfield  {author} {\bibinfo {author} {\bibfnamefont {Y.}~\bibnamefont
  {Iqbal}}, \bibinfo {author} {\bibfnamefont {F.}~\bibnamefont {Becca}},
  \bibinfo {author} {\bibfnamefont {S.}~\bibnamefont {Sorella}},\ and\ \bibinfo
  {author} {\bibfnamefont {D.}~\bibnamefont {Poilblanc}},\ }\bibfield  {title}
  {\bibinfo {title} {Gapless spin-liquid phase in the kagome spin-$\frac{1}{2}$
  heisenberg antiferromagnet},\ }\href
  {https://doi.org/10.1103/PhysRevB.87.060405} {\bibfield  {journal} {\bibinfo
  {journal} {Phys. Rev. B}\ }\textbf {\bibinfo {volume} {87}},\ \bibinfo
  {pages} {060405} (\bibinfo {year} {2013})}\BibitemShut {NoStop}%
\bibitem [{\citenamefont {Depenbrock}\ \emph {et~al.}(2012)\citenamefont
  {Depenbrock}, \citenamefont {McCulloch},\ and\ \citenamefont
  {Schollw\"ock}}]{prl067201}%
  \BibitemOpen
  \bibfield  {author} {\bibinfo {author} {\bibfnamefont {S.}~\bibnamefont
  {Depenbrock}}, \bibinfo {author} {\bibfnamefont {I.~P.}\ \bibnamefont
  {McCulloch}},\ and\ \bibinfo {author} {\bibfnamefont {U.}~\bibnamefont
  {Schollw\"ock}},\ }\bibfield  {title} {\bibinfo {title} {Nature of the
  spin-liquid ground state of the $s=1/2$ heisenberg model on the kagome
  lattice},\ }\href {https://doi.org/10.1103/PhysRevLett.109.067201} {\bibfield
   {journal} {\bibinfo  {journal} {Phys. Rev. Lett.}\ }\textbf {\bibinfo
  {volume} {109}},\ \bibinfo {pages} {067201} (\bibinfo {year}
  {2012})}\BibitemShut {NoStop}%
\bibitem [{\citenamefont {Yu}\ and\ \citenamefont
  {Li}(2012)}]{PhysRevB.85.144402}%
  \BibitemOpen
  \bibfield  {author} {\bibinfo {author} {\bibfnamefont {S.-L.}\ \bibnamefont
  {Yu}}\ and\ \bibinfo {author} {\bibfnamefont {J.-X.}\ \bibnamefont {Li}},\
  }\bibfield  {title} {\bibinfo {title} {Chiral superconducting phase and
  chiral spin-density-wave phase in a hubbard model on the kagome lattice},\
  }\href {https://doi.org/10.1103/PhysRevB.85.144402} {\bibfield  {journal}
  {\bibinfo  {journal} {Phys. Rev. B}\ }\textbf {\bibinfo {volume} {85}},\
  \bibinfo {pages} {144402} (\bibinfo {year} {2012})}\BibitemShut {NoStop}%
\bibitem [{\citenamefont {Wang}\ \emph {et~al.}(2013)\citenamefont {Wang},
  \citenamefont {Li}, \citenamefont {Xiang},\ and\ \citenamefont
  {Wang}}]{PhysRevB.87.115135}%
  \BibitemOpen
  \bibfield  {author} {\bibinfo {author} {\bibfnamefont {W.-S.}\ \bibnamefont
  {Wang}}, \bibinfo {author} {\bibfnamefont {Z.-Z.}\ \bibnamefont {Li}},
  \bibinfo {author} {\bibfnamefont {Y.-Y.}\ \bibnamefont {Xiang}},\ and\
  \bibinfo {author} {\bibfnamefont {Q.-H.}\ \bibnamefont {Wang}},\ }\bibfield
  {title} {\bibinfo {title} {Competing electronic orders on kagome lattices at
  van hove filling},\ }\href {https://doi.org/10.1103/PhysRevB.87.115135}
  {\bibfield  {journal} {\bibinfo  {journal} {Phys. Rev. B}\ }\textbf {\bibinfo
  {volume} {87}},\ \bibinfo {pages} {115135} (\bibinfo {year}
  {2013})}\BibitemShut {NoStop}%
\bibitem [{\citenamefont {Kiesel}\ \emph {et~al.}(2013)\citenamefont {Kiesel},
  \citenamefont {Platt},\ and\ \citenamefont {Thomale}}]{Kiesel2013}%
  \BibitemOpen
  \bibfield  {author} {\bibinfo {author} {\bibfnamefont {M.~L.}\ \bibnamefont
  {Kiesel}}, \bibinfo {author} {\bibfnamefont {C.}~\bibnamefont {Platt}},\ and\
  \bibinfo {author} {\bibfnamefont {R.}~\bibnamefont {Thomale}},\ }\bibfield
  {title} {\bibinfo {title} {Unconventional fermi surface instabilities in the
  kagome hubbard model},\ }\href
  {https://doi.org/10.1103/PhysRevLett.110.126405} {\bibfield  {journal}
  {\bibinfo  {journal} {Phys. Rev. Lett.}\ }\textbf {\bibinfo {volume} {110}},\
  \bibinfo {pages} {126405} (\bibinfo {year} {2013})}\BibitemShut {NoStop}%
\bibitem [{\citenamefont {Chen}\ \emph {et~al.}(2019)\citenamefont {Chen},
  \citenamefont {Liu},\ and\ \citenamefont {Zheng}}]{PhysRevB.99.085119}%
  \BibitemOpen
  \bibfield  {author} {\bibinfo {author} {\bibfnamefont {L.-H.}\ \bibnamefont
  {Chen}}, \bibinfo {author} {\bibfnamefont {Z.}~\bibnamefont {Liu}},\ and\
  \bibinfo {author} {\bibfnamefont {J.-T.}\ \bibnamefont {Zheng}},\ }\bibfield
  {title} {\bibinfo {title} {Matrix element interference in $n$-patch
  functional renormalization group},\ }\href
  {https://doi.org/10.1103/PhysRevB.99.085119} {\bibfield  {journal} {\bibinfo
  {journal} {Phys. Rev. B}\ }\textbf {\bibinfo {volume} {99}},\ \bibinfo
  {pages} {085119} (\bibinfo {year} {2019})}\BibitemShut {NoStop}%
\bibitem [{\citenamefont {Ortiz}\ \emph {et~al.}(2019)\citenamefont {Ortiz},
  \citenamefont {Gomes}, \citenamefont {Morey}, \citenamefont {Winiarski},
  \citenamefont {Bordelon}, \citenamefont {Mangum}, \citenamefont {Oswald},
  \citenamefont {Rodriguez-Rivera}, \citenamefont {Neilson}, \citenamefont
  {Wilson}, \citenamefont {Ertekin}, \citenamefont {McQueen},\ and\
  \citenamefont {Toberer}}]{kagome.material}%
  \BibitemOpen
  \bibfield  {author} {\bibinfo {author} {\bibfnamefont {B.~R.}\ \bibnamefont
  {Ortiz}}, \bibinfo {author} {\bibfnamefont {L.~C.}\ \bibnamefont {Gomes}},
  \bibinfo {author} {\bibfnamefont {J.~R.}\ \bibnamefont {Morey}}, \bibinfo
  {author} {\bibfnamefont {M.}~\bibnamefont {Winiarski}}, \bibinfo {author}
  {\bibfnamefont {M.}~\bibnamefont {Bordelon}}, \bibinfo {author}
  {\bibfnamefont {J.~S.}\ \bibnamefont {Mangum}}, \bibinfo {author}
  {\bibfnamefont {I.~W.~H.}\ \bibnamefont {Oswald}}, \bibinfo {author}
  {\bibfnamefont {J.~A.}\ \bibnamefont {Rodriguez-Rivera}}, \bibinfo {author}
  {\bibfnamefont {J.~R.}\ \bibnamefont {Neilson}}, \bibinfo {author}
  {\bibfnamefont {S.~D.}\ \bibnamefont {Wilson}}, \bibinfo {author}
  {\bibfnamefont {E.}~\bibnamefont {Ertekin}}, \bibinfo {author} {\bibfnamefont
  {T.~M.}\ \bibnamefont {McQueen}},\ and\ \bibinfo {author} {\bibfnamefont
  {E.~S.}\ \bibnamefont {Toberer}},\ }\bibfield  {title} {\bibinfo {title} {New
  kagome prototype materials: discovery of
  {${\mathrm{KV}}_{3}{\mathrm{Sb}}_{5},{\mathrm{RbV}}_{3}{\mathrm{Sb}}_{5}$,
  and ${\mathrm{CsV}}_{3}{\mathrm{Sb}}_{5}$}},\ }\href
  {https://doi.org/10.1103/PhysRevMaterials.3.094407} {\bibfield  {journal}
  {\bibinfo  {journal} {Phys. Rev. Mater.}\ }\textbf {\bibinfo {volume} {3}},\
  \bibinfo {pages} {094407} (\bibinfo {year} {2019})}\BibitemShut {NoStop}%
\bibitem [{\citenamefont {Ortiz}\ \emph {et~al.}(2020)\citenamefont {Ortiz},
  \citenamefont {Teicher}, \citenamefont {Hu}, \citenamefont {Zuo},
  \citenamefont {Sarte}, \citenamefont {Schueller}, \citenamefont {Abeykoon},
  \citenamefont {Krogstad}, \citenamefont {Rosenkranz}, \citenamefont {Osborn},
  \citenamefont {Seshadri}, \citenamefont {Balents}, \citenamefont {He},\ and\
  \citenamefont {Wilson}}]{PhysRevLett.125.247002}%
  \BibitemOpen
  \bibfield  {author} {\bibinfo {author} {\bibfnamefont {B.~R.}\ \bibnamefont
  {Ortiz}}, \bibinfo {author} {\bibfnamefont {S.~M.~L.}\ \bibnamefont
  {Teicher}}, \bibinfo {author} {\bibfnamefont {Y.}~\bibnamefont {Hu}},
  \bibinfo {author} {\bibfnamefont {J.~L.}\ \bibnamefont {Zuo}}, \bibinfo
  {author} {\bibfnamefont {P.~M.}\ \bibnamefont {Sarte}}, \bibinfo {author}
  {\bibfnamefont {E.~C.}\ \bibnamefont {Schueller}}, \bibinfo {author}
  {\bibfnamefont {A.~M.~M.}\ \bibnamefont {Abeykoon}}, \bibinfo {author}
  {\bibfnamefont {M.~J.}\ \bibnamefont {Krogstad}}, \bibinfo {author}
  {\bibfnamefont {S.}~\bibnamefont {Rosenkranz}}, \bibinfo {author}
  {\bibfnamefont {R.}~\bibnamefont {Osborn}}, \bibinfo {author} {\bibfnamefont
  {R.}~\bibnamefont {Seshadri}}, \bibinfo {author} {\bibfnamefont
  {L.}~\bibnamefont {Balents}}, \bibinfo {author} {\bibfnamefont
  {J.}~\bibnamefont {He}},\ and\ \bibinfo {author} {\bibfnamefont {S.~D.}\
  \bibnamefont {Wilson}},\ }\bibfield  {title} {\bibinfo {title}
  {{$\mathrm{Cs}{\mathrm{V}}_{3}{\mathrm{Sb}}_{5}$: A ${\mathbb{Z}}_{2}$
  Topological Kagome Metal with a Superconducting Ground State}},\ }\href
  {https://doi.org/10.1103/PhysRevLett.125.247002} {\bibfield  {journal}
  {\bibinfo  {journal} {Phys. Rev. Lett.}\ }\textbf {\bibinfo {volume} {125}},\
  \bibinfo {pages} {247002} (\bibinfo {year} {2020})}\BibitemShut {NoStop}%
\bibitem [{\citenamefont {Ortiz}\ \emph {et~al.}(2021)\citenamefont {Ortiz},
  \citenamefont {Sarte}, \citenamefont {Kenney}, \citenamefont {Graf},
  \citenamefont {Teicher}, \citenamefont {Seshadri},\ and\ \citenamefont
  {Wilson}}]{PhysRevMaterials.5.034801}%
  \BibitemOpen
  \bibfield  {author} {\bibinfo {author} {\bibfnamefont {B.~R.}\ \bibnamefont
  {Ortiz}}, \bibinfo {author} {\bibfnamefont {P.~M.}\ \bibnamefont {Sarte}},
  \bibinfo {author} {\bibfnamefont {E.~M.}\ \bibnamefont {Kenney}}, \bibinfo
  {author} {\bibfnamefont {M.~J.}\ \bibnamefont {Graf}}, \bibinfo {author}
  {\bibfnamefont {S.~M.~L.}\ \bibnamefont {Teicher}}, \bibinfo {author}
  {\bibfnamefont {R.}~\bibnamefont {Seshadri}},\ and\ \bibinfo {author}
  {\bibfnamefont {S.~D.}\ \bibnamefont {Wilson}},\ }\bibfield  {title}
  {\bibinfo {title} {{Superconductivity in the ${\mathbb{Z}}_{2}$ kagome metal
  ${\mathrm{KV}}_{3}{\mathrm{Sb}}_{5}$}},\ }\href
  {https://doi.org/10.1103/PhysRevMaterials.5.034801} {\bibfield  {journal}
  {\bibinfo  {journal} {Phys. Rev. Mater.}\ }\textbf {\bibinfo {volume} {5}},\
  \bibinfo {pages} {034801} (\bibinfo {year} {2021})}\BibitemShut {NoStop}%
\bibitem [{\citenamefont {Yin}\ \emph {et~al.}(2021)\citenamefont {Yin},
  \citenamefont {Tu}, \citenamefont {Gong}, \citenamefont {Fu}, \citenamefont
  {Yan},\ and\ \citenamefont {Lei}}]{Yin_2021}%
  \BibitemOpen
  \bibfield  {author} {\bibinfo {author} {\bibfnamefont {Q.}~\bibnamefont
  {Yin}}, \bibinfo {author} {\bibfnamefont {Z.}~\bibnamefont {Tu}}, \bibinfo
  {author} {\bibfnamefont {C.}~\bibnamefont {Gong}}, \bibinfo {author}
  {\bibfnamefont {Y.}~\bibnamefont {Fu}}, \bibinfo {author} {\bibfnamefont
  {S.}~\bibnamefont {Yan}},\ and\ \bibinfo {author} {\bibfnamefont
  {H.}~\bibnamefont {Lei}},\ }\bibfield  {title} {\bibinfo {title}
  {Superconductivity and normal-state properties of kagome metal rbv3sb5 single
  crystals},\ }\href {https://doi.org/10.1088/0256-307X/38/3/037403} {\bibfield
   {journal} {\bibinfo  {journal} {Chin. Phys. Lett.}\ }\textbf {\bibinfo
  {volume} {38}},\ \bibinfo {pages} {037403} (\bibinfo {year}
  {2021})}\BibitemShut {NoStop}%
\bibitem [{\citenamefont {Barros}\ \emph {et~al.}(2014)\citenamefont {Barros},
  \citenamefont {Venderbos}, \citenamefont {Chern},\ and\ \citenamefont
  {Batista}}]{PhysRevB.90.245119}%
  \BibitemOpen
  \bibfield  {author} {\bibinfo {author} {\bibfnamefont {K.}~\bibnamefont
  {Barros}}, \bibinfo {author} {\bibfnamefont {J.~W.~F.}\ \bibnamefont
  {Venderbos}}, \bibinfo {author} {\bibfnamefont {G.-W.}\ \bibnamefont
  {Chern}},\ and\ \bibinfo {author} {\bibfnamefont {C.~D.}\ \bibnamefont
  {Batista}},\ }\bibfield  {title} {\bibinfo {title} {Exotic magnetic orderings
  in the kagome kondo-lattice model},\ }\href
  {https://doi.org/10.1103/PhysRevB.90.245119} {\bibfield  {journal} {\bibinfo
  {journal} {Phys. Rev. B}\ }\textbf {\bibinfo {volume} {90}},\ \bibinfo
  {pages} {245119} (\bibinfo {year} {2014})}\BibitemShut {NoStop}%
\bibitem [{\citenamefont {Kiesel}\ and\ \citenamefont
  {Thomale}(2012)}]{PhysRevB.86.121105}%
  \BibitemOpen
  \bibfield  {author} {\bibinfo {author} {\bibfnamefont {M.~L.}\ \bibnamefont
  {Kiesel}}\ and\ \bibinfo {author} {\bibfnamefont {R.}~\bibnamefont
  {Thomale}},\ }\bibfield  {title} {\bibinfo {title} {Sublattice interference
  in the kagome hubbard model},\ }\href
  {https://doi.org/10.1103/PhysRevB.86.121105} {\bibfield  {journal} {\bibinfo
  {journal} {Phys. Rev. B}\ }\textbf {\bibinfo {volume} {86}},\ \bibinfo
  {pages} {121105} (\bibinfo {year} {2012})}\BibitemShut {NoStop}%
\bibitem [{\citenamefont {van Loon}\ \emph {et~al.}(2016)\citenamefont {van
  Loon}, \citenamefont {Sch\"uler}, \citenamefont {Katsnelson},\ and\
  \citenamefont {Wehling}}]{PhysRevB.94.165141}%
  \BibitemOpen
  \bibfield  {author} {\bibinfo {author} {\bibfnamefont {E.~G. C.~P.}\
  \bibnamefont {van Loon}}, \bibinfo {author} {\bibfnamefont {M.}~\bibnamefont
  {Sch\"uler}}, \bibinfo {author} {\bibfnamefont {M.~I.}\ \bibnamefont
  {Katsnelson}},\ and\ \bibinfo {author} {\bibfnamefont {T.~O.}\ \bibnamefont
  {Wehling}},\ }\bibfield  {title} {\bibinfo {title} {Capturing nonlocal
  interaction effects in the hubbard model: Optimal mappings and limits of
  applicability},\ }\href {https://doi.org/10.1103/PhysRevB.94.165141}
  {\bibfield  {journal} {\bibinfo  {journal} {Phys. Rev. B}\ }\textbf {\bibinfo
  {volume} {94}},\ \bibinfo {pages} {165141} (\bibinfo {year}
  {2016})}\BibitemShut {NoStop}%
\bibitem [{\citenamefont {Wang}\ \emph {et~al.}(2012)\citenamefont {Wang},
  \citenamefont {Xiang}, \citenamefont {Wang}, \citenamefont {Wang},
  \citenamefont {Yang},\ and\ \citenamefont {Lee}}]{Wang_PRB_2012}%
  \BibitemOpen
  \bibfield  {author} {\bibinfo {author} {\bibfnamefont {W.-S.}\ \bibnamefont
  {Wang}}, \bibinfo {author} {\bibfnamefont {Y.-Y.}\ \bibnamefont {Xiang}},
  \bibinfo {author} {\bibfnamefont {Q.-H.}\ \bibnamefont {Wang}}, \bibinfo
  {author} {\bibfnamefont {F.}~\bibnamefont {Wang}}, \bibinfo {author}
  {\bibfnamefont {F.}~\bibnamefont {Yang}},\ and\ \bibinfo {author}
  {\bibfnamefont {D.-H.}\ \bibnamefont {Lee}},\ }\bibfield  {title} {\bibinfo
  {title} {Functional renormalization group and variational {{Monte Carlo}}
  studies of the electronic instabilities in graphene near 1/4 doping},\ }\href
  {https://doi.org/10.1103/PhysRevB.85.035414} {\bibfield  {journal} {\bibinfo
  {journal} {Phys. Rev. B}\ }\textbf {\bibinfo {volume} {85}},\ \bibinfo
  {pages} {035414} (\bibinfo {year} {2012})}\BibitemShut {NoStop}%
\bibitem [{\citenamefont {Husemann}\ and\ \citenamefont
  {Salmhofer}(2009)}]{PhysRevB.79.195125}%
  \BibitemOpen
  \bibfield  {author} {\bibinfo {author} {\bibfnamefont {C.}~\bibnamefont
  {Husemann}}\ and\ \bibinfo {author} {\bibfnamefont {M.}~\bibnamefont
  {Salmhofer}},\ }\bibfield  {title} {\bibinfo {title} {Efficient
  parametrization of the vertex function, $\ensuremath{\Omega}$ scheme, and the
  $t,{t}^{\ensuremath{'}}$ hubbard model at van hove filling},\ }\href
  {https://doi.org/10.1103/PhysRevB.79.195125} {\bibfield  {journal} {\bibinfo
  {journal} {Phys. Rev. B}\ }\textbf {\bibinfo {volume} {79}},\ \bibinfo
  {pages} {195125} (\bibinfo {year} {2009})}\BibitemShut {NoStop}%
\bibitem [{\citenamefont {Xiang}\ \emph {et~al.}(2012)\citenamefont {Xiang},
  \citenamefont {Wang}, \citenamefont {Wang},\ and\ \citenamefont
  {Lee}}]{Xiang_PRB_2012}%
  \BibitemOpen
  \bibfield  {author} {\bibinfo {author} {\bibfnamefont {Y.-Y.}\ \bibnamefont
  {Xiang}}, \bibinfo {author} {\bibfnamefont {W.-S.}\ \bibnamefont {Wang}},
  \bibinfo {author} {\bibfnamefont {Q.-H.}\ \bibnamefont {Wang}},\ and\
  \bibinfo {author} {\bibfnamefont {D.-H.}\ \bibnamefont {Lee}},\ }\bibfield
  {title} {\bibinfo {title} {Topological superconducting phase in the vicinity
  of ferromagnetic phases},\ }\href
  {https://doi.org/10.1103/PhysRevB.86.024523} {\bibfield  {journal} {\bibinfo
  {journal} {Phys. Rev. B}\ }\textbf {\bibinfo {volume} {86}},\ \bibinfo
  {pages} {024523} (\bibinfo {year} {2012})}\BibitemShut {NoStop}%
\bibitem [{\citenamefont {Wang}\ \emph {et~al.}(2019)\citenamefont {Wang},
  \citenamefont {Zhang}, \citenamefont {Zhang},\ and\ \citenamefont
  {Wang}}]{Wang_PRL_2019}%
  \BibitemOpen
  \bibfield  {author} {\bibinfo {author} {\bibfnamefont {W.-S.}\ \bibnamefont
  {Wang}}, \bibinfo {author} {\bibfnamefont {C.-C.}\ \bibnamefont {Zhang}},
  \bibinfo {author} {\bibfnamefont {F.-C.}\ \bibnamefont {Zhang}},\ and\
  \bibinfo {author} {\bibfnamefont {Q.-H.}\ \bibnamefont {Wang}},\ }\bibfield
  {title} {\bibinfo {title} {Theory of {{Chiral}} $p$-{{Wave
  Superconductivity}} with {{Near Nodes}} for {Sr2RuO4}},\ }\href
  {https://doi.org/10.1103/PhysRevLett.122.027002} {\bibfield  {journal}
  {\bibinfo  {journal} {Phys. Rev. Lett.}\ }\textbf {\bibinfo {volume} {122}},\
  \bibinfo {pages} {027002} (\bibinfo {year} {2019})}\BibitemShut {NoStop}%
\bibitem [{\citenamefont {Yang}\ \emph {et~al.}(2022)\citenamefont {Yang},
  \citenamefont {Wang},\ and\ \citenamefont {Wang}}]{Yang_PRB_2022}%
  \BibitemOpen
  \bibfield  {author} {\bibinfo {author} {\bibfnamefont {Q.-G.}\ \bibnamefont
  {Yang}}, \bibinfo {author} {\bibfnamefont {D.}~\bibnamefont {Wang}},\ and\
  \bibinfo {author} {\bibfnamefont {Q.-H.}\ \bibnamefont {Wang}},\ }\bibfield
  {title} {\bibinfo {title} {Functional renormalization group study of the
  two-dimensional {Su}-{Schrieffer}-{Heeger}-{Hubbard} model},\ }\href
  {https://doi.org/10.1103/PhysRevB.106.245136} {\bibfield  {journal} {\bibinfo
   {journal} {Phys. Rev. B}\ }\textbf {\bibinfo {volume} {106}},\ \bibinfo
  {pages} {245136} (\bibinfo {year} {2022})}\BibitemShut {NoStop}%
\bibitem [{\citenamefont {Yang}\ \emph {et~al.}(2023)\citenamefont {Yang},
  \citenamefont {Wang},\ and\ \citenamefont {Wang}}]{Yang_PRB_2023}%
  \BibitemOpen
  \bibfield  {author} {\bibinfo {author} {\bibfnamefont {Q.-G.}\ \bibnamefont
  {Yang}}, \bibinfo {author} {\bibfnamefont {D.}~\bibnamefont {Wang}},\ and\
  \bibinfo {author} {\bibfnamefont {Q.-H.}\ \bibnamefont {Wang}},\ }\bibfield
  {title} {\bibinfo {title} {Possible {$S_\pm$}-wave superconductivity in
  {La$_3$Ni$_2$O$_7$}},\ }\href {https://doi.org/10.1103/PhysRevB.108.L140505}
  {\bibfield  {journal} {\bibinfo  {journal} {Phys. Rev. B}\ }\textbf {\bibinfo
  {volume} {108}},\ \bibinfo {pages} {L140505} (\bibinfo {year}
  {2023})}\BibitemShut {NoStop}%
\bibitem [{\citenamefont {Becca}\ and\ \citenamefont
  {Sorella}(2017)}]{vmcbook}%
  \BibitemOpen
  \bibfield  {author} {\bibinfo {author} {\bibfnamefont {F.}~\bibnamefont
  {Becca}}\ and\ \bibinfo {author} {\bibfnamefont {S.}~\bibnamefont
  {Sorella}},\ }\href {https://doi.org/10.1017/9781316417041} {\emph {\bibinfo
  {title} {Quantum Monte Carlo Approaches for Correlated Systems}}}\ (\bibinfo
  {publisher} {Cambridge University Press},\ \bibinfo {year} {2017})\ pp.\
  \bibinfo {pages} {103--129}\BibitemShut {NoStop}%
\bibitem [{\citenamefont {Kolorenč}\ and\ \citenamefont
  {Mitas}(2011)}]{vmcreview1}%
  \BibitemOpen
  \bibfield  {author} {\bibinfo {author} {\bibfnamefont {J.}~\bibnamefont
  {Kolorenč}}\ and\ \bibinfo {author} {\bibfnamefont {L.}~\bibnamefont
  {Mitas}},\ }\bibfield  {title} {\bibinfo {title} {Applications of quantum
  monte carlo methods in condensed systems},\ }\href
  {https://doi.org/10.1088/0034-4885/74/2/026502} {\bibfield  {journal}
  {\bibinfo  {journal} {Rep. Prog. Phys.}\ }\textbf {\bibinfo {volume} {74}},\
  \bibinfo {pages} {026502} (\bibinfo {year} {2011})}\BibitemShut {NoStop}%
\bibitem [{\citenamefont {Pollmann}\ \emph {et~al.}(2008)\citenamefont
  {Pollmann}, \citenamefont {Fulde},\ and\ \citenamefont
  {Shtengel}}]{PhysRevLett.100.136404}%
  \BibitemOpen
  \bibfield  {author} {\bibinfo {author} {\bibfnamefont {F.}~\bibnamefont
  {Pollmann}}, \bibinfo {author} {\bibfnamefont {P.}~\bibnamefont {Fulde}},\
  and\ \bibinfo {author} {\bibfnamefont {K.}~\bibnamefont {Shtengel}},\
  }\bibfield  {title} {\bibinfo {title} {Kinetic ferromagnetism on a kagome
  lattice},\ }\href {https://doi.org/10.1103/PhysRevLett.100.136404} {\bibfield
   {journal} {\bibinfo  {journal} {Phys. Rev. Lett.}\ }\textbf {\bibinfo
  {volume} {100}},\ \bibinfo {pages} {136404} (\bibinfo {year}
  {2008})}\BibitemShut {NoStop}%
\end{thebibliography}%
\end{document}